\documentclass[aps,twocolumn,superscriptaddress,nofootinbib]{revtex4-1}

\usepackage{graphicx,enumerate,verbatim,bbold}
\usepackage{amsmath,amssymb,amsthm,float,mathrsfs}
\usepackage{dsfont}
\usepackage[colorlinks]{hyperref}
\usepackage[T1]{fontenc}
\usepackage[utf8]{inputenc}
\usepackage{lmodern}
\usepackage{braket}
\usepackage[caption=false]{subfig}
\usepackage{dcolumn}

\begin{document}

\title{Rabi oscillation in a quantum cavity: Markovian and non-Markovian dynamics}

\author{Pierre-Olivier Guimond}
\affiliation{Centre for Quantum Technologies, National University of Singapore, 3 Science Drive 2, Singapore 117543, Singapore}
\affiliation{Ecole Polytechnique Fédérale de Lausanne (EPFL), CH-1015 Lausanne, Switzerland}

\author{Alexandre Roulet}
\affiliation{Centre for Quantum Technologies, National University of Singapore, 3 Science Drive 2, Singapore 117543, Singapore}

\author{Huy Nguyen Le}
\affiliation{Centre for Quantum Technologies, National University of Singapore, 3 Science Drive 2, Singapore 117543, Singapore}

\author{Valerio Scarani}
\affiliation{Centre for Quantum Technologies, National University of Singapore, 3 Science Drive 2, Singapore 117543, Singapore}
\affiliation{Department of Physics, National University of Singapore, 2 Science Drive 3, Singapore 117542, Singapore}

\date{\today}

\begin{abstract}
We investigate the Rabi oscillation of an atom placed inside a quantum cavity where each mirror is formed by a chain of atoms trapped near a one-dimensional waveguide. This proposal was studied previously with the use of Markov approximation, where the delay due to the finite travel time of light between the two cavity mirrors is neglected. We show that Rabi oscillation analogous to that obtained with high-finesse classical cavities is achieved only when this travel time is much larger than the time scale that characterizes the superradiant response of the mirrors. Therefore, the delay must be taken into account and the dynamics of the problem is inherently non-Markovian. Parameters of interest such as the Rabi frequency and the cavity loss rate due to photon leakage through the mirrors are obtained.
\end{abstract}


\maketitle

\section{Introduction}
Many applications in quantum information processing rely on the strong interaction between stationary two-level emitters and photons~\cite{opticalqip2011}. This can be realized in cavity-QED (CQED) where an atom is placed inside a high finesse cavity~\cite{Turchette1995,Domokos1995,Imamoglu1999,Zheng2000,Blais2004}, or waveguide-QED where the atom is placed inside or near a one-dimensional (1D) waveguide such as hollow-core fibers, fiber-taper waveguides and photonic crystal waveguides~\cite{Christensen2008,Bajcsy2011,Vetsch2010,Goban2012,Arcari2014,Goban2014}. As CQED has become a well-established field, important experimental progress has been made toward enhanced coupling between atoms and the field in 1D waveguides~\cite{Arcari2014,Goban2014}. 

\begin{figure}
\subfloat{
\includegraphics[width=0.43\textwidth]{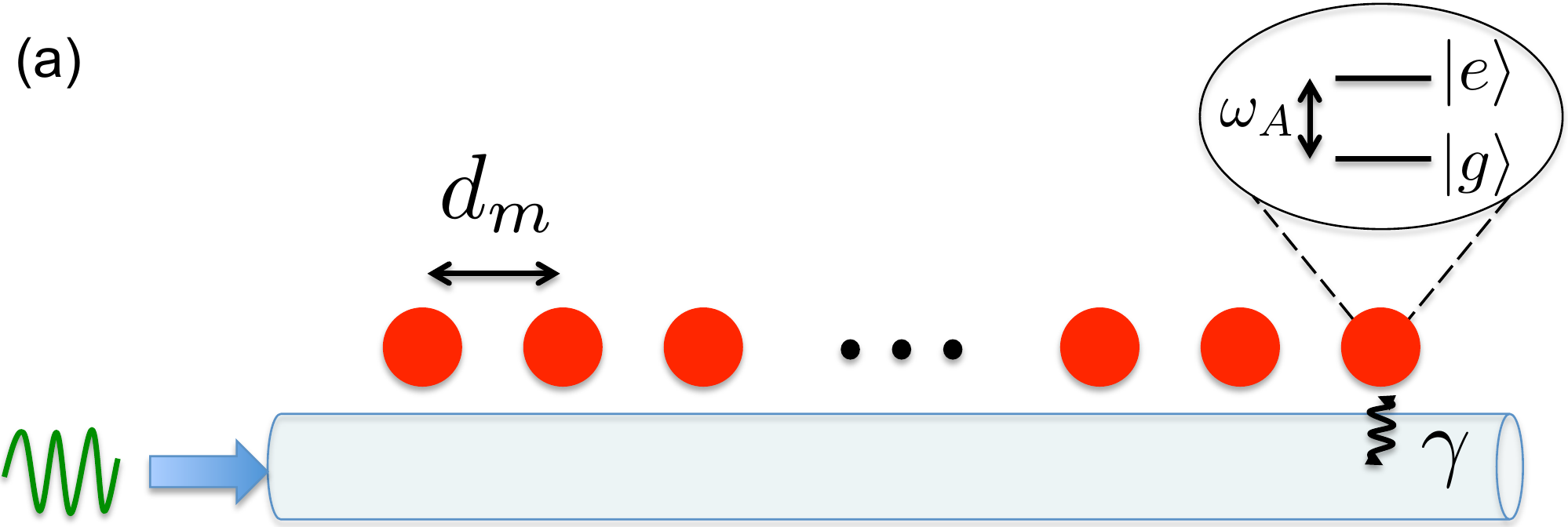}\label{fig:mirror}
}\\
\subfloat{
\includegraphics[width=0.43\textwidth]{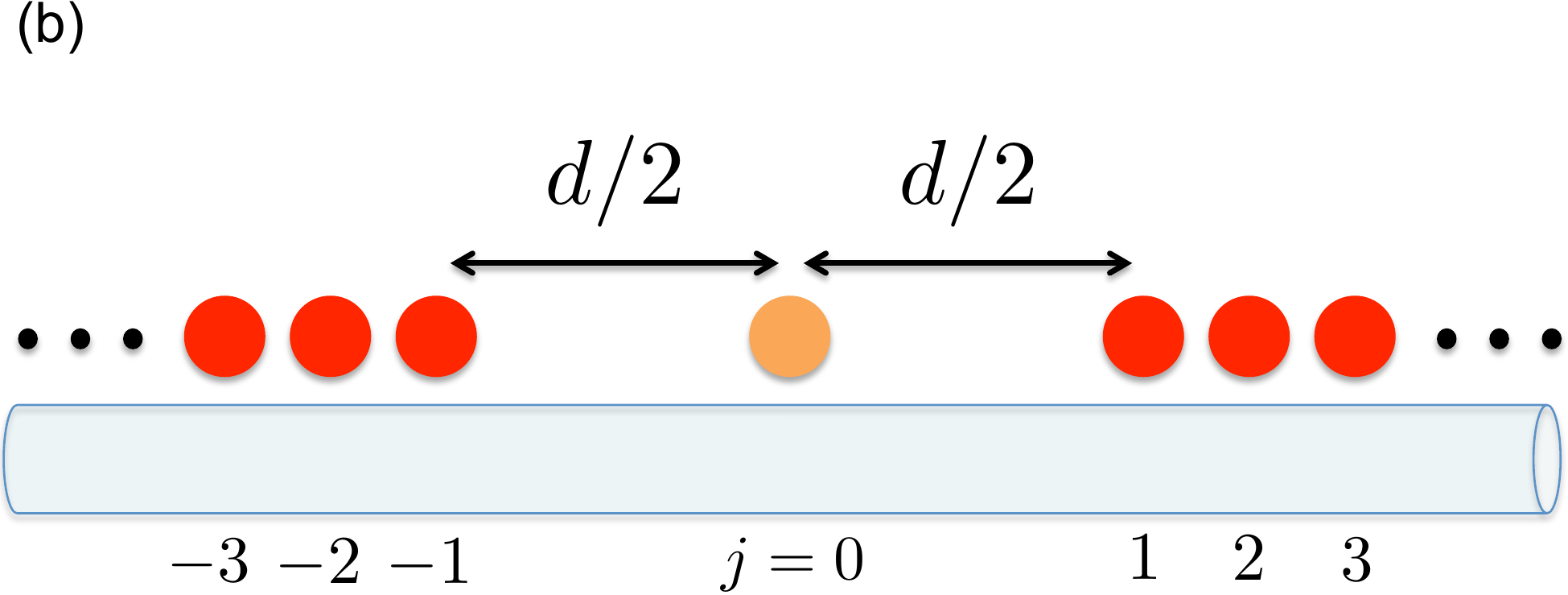}\label{fig:cavity}
}
\caption{\label{fig:Qcavity}(color online).\quad Cavity QED using atomic mirrors. \\(a) Scattering of an incoming photon on a chain of atoms with lattice constant $d_m$. When $\omega_A d_m/v_g=l\pi$ with $l$ an integer, the chain forms a Bragg mirror.  (b) An initially excited atom (orange) is sitting inside a quantum cavity formed by two atomic mirrors (red). The spatial extension of the mirrors is small compared to the cavity length $d$.}
\end{figure}

A hybrid strategy that combines the appealing features of both approaches has been proposed by Chang~\emph{et~al.} in Ref.\,\cite{Chang2012}, where a chain of atoms trapped near a 1D waveguide is used as a mirror in a cavity setup (see Fig.\,\ref{fig:Qcavity}). In this seminal work, it is assumed that the collective response time of each atomic mirror is much longer than the time it takes for the photon to travel from one mirror to the other, and as a consequence the delay due to this travel time is neglected. This approach is the Markov approximation commonly used to accurately describe the interaction of photons with many atoms in a wide range of experimental situations. There are, however, certain cases where this assumption is not necessarily justified \cite{Zheng2013,Shi2015}. Here, we show that in the Markovian regime the lifetime of the photon inside the cavity is not enhanced by the presence of the mirrors, and therefore sustained Rabi oscillation analogous to that observed in conventional CQED setups cannot be obtained. Indeed, it is in the non-Markovian regime, where the response time of the atomic mirrors is much smaller than the delay, that Rabi oscillation in the usual sense is achieved.

As depicted in Fig.\,\ref{fig:Qcavity}, the problem under consideration involves an atom, initially in the excited state, located between two quantum mirrors each consisting of $N$ equally spaced atoms trapped near a 1D waveguide. The position of this central atom is at an antinode of the standing wave in the cavity in order to maximize the coupling. We study how the excitation probability amplitude of the central atom $c_0(t)$ evolves with time. Our approach is valid for both Markovian and non-Markovian regimes. The figures of merit such as the Rabi frequency and the cavity loss rate through the mirrors are computed. Our main result is that the vacuum Rabi oscillation of the central atom is given by
\begin{equation}
c_0(t)\approx e^{-\frac{\gamma t}{2(1+N\gamma d/v_g)^2}}\, \cos{\left[\sqrt{\frac{2N}{1+N \gamma d/v_g}}\gamma t\right]},
\end{equation}
where $d$ is the distance between the two atomic mirrors, $\gamma$ the single-atom decay rate into the waveguide modes and $v_g$ the group velocity of light in the cavity.  One sees from the above expression that the cavity loss rate is $\kappa=\gamma/(1+N\gamma d/v_g)^2$ and the Rabi frequency is  $\Omega_{\text{Rabi}}~=~\gamma\sqrt{2N/(1+N \gamma d/v_g)}$. 

The results obtained in Ref.\,\cite{Chang2012}, where $\Omega_{\text{Rabi}}\approx \gamma\sqrt{2N} $ and the loss rate is comparable to $\gamma$, are retrieved in the limit $N\gamma d/v_g\rightarrow 0$, which corresponds to the Markov approximation. At the other extreme when $ d/v_g \gg 1/(N \gamma)$, the lifetime of the photon inside the cavity is greatly enhanced similarly to conventional CQED setups. Since $N \gamma$ is the collective decay rate of the atomic mirror, $1/(N \gamma)$ is the time scale of the mirror's response. Thus, the strong interaction regime analogous to that achieved with high-finesse classical mirrors is inherently non-Markovian. In this regime, the Rabi frequency $\Omega_{\text{Rabi}}\approx\gamma\sqrt{2v_g/(\gamma d)}$ depends on the cavity length instead of the number of atoms in the mirrors. The $\sim1/\sqrt{d}$ dependence is expected because in one dimension $d$ plays the role of the modal volume of the field in the cavity.

\section{Quantum cavity}
\begin{figure}
\includegraphics[width=0.46\textwidth]{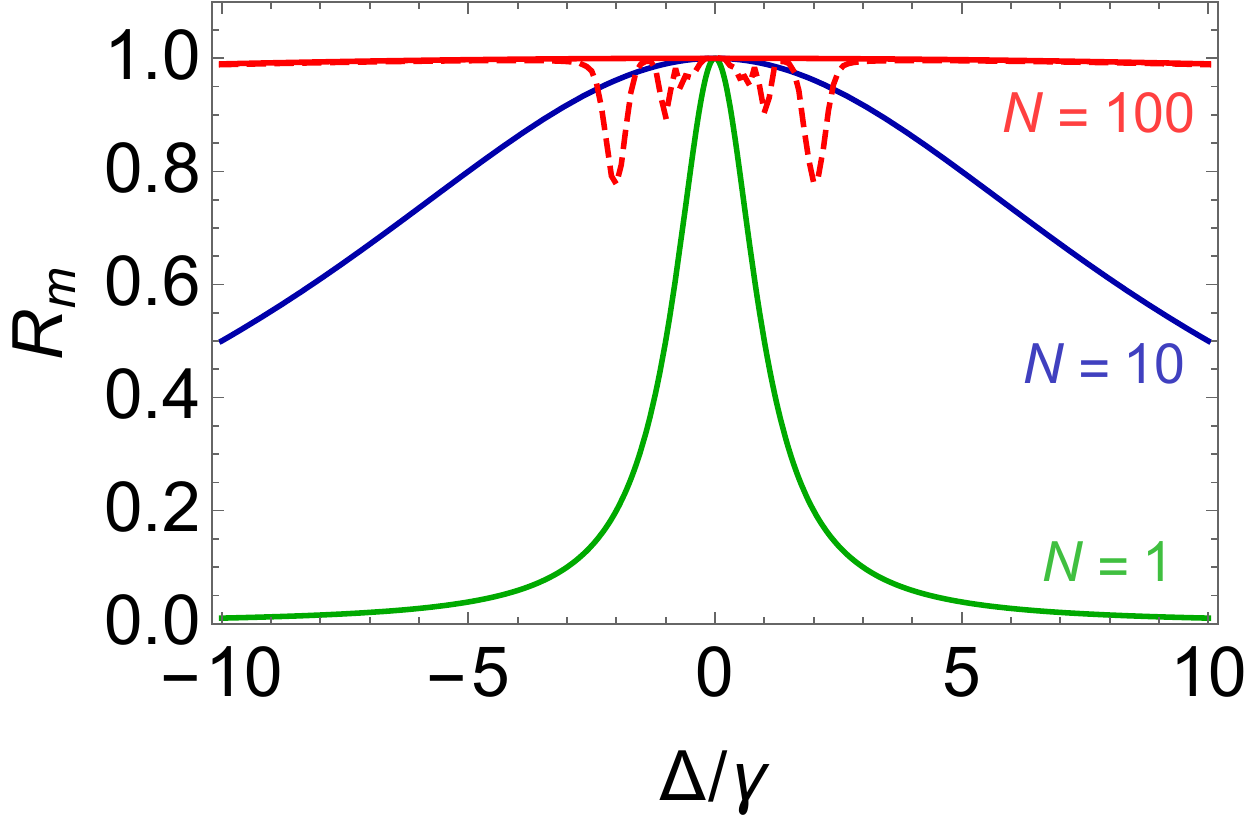}
\caption{\label{fig:reflectance}(color online).\quad Reflectance of an atomic mirror as a function of the normalized detuning $\Delta/\gamma$. The green, blue and red line represent respectively a mirror made of $N=1,10, \text{ and } 100$ atoms. For reference, the dashed line illustrates the case where each atom forming the mirror fluctuates around its average position $x_j$ with $\sqrt{\langle \delta x_j^2 \rangle}= 0.01 d_m$.  $\omega_A d_m/c=\pi$.}
\end{figure}

Let us first describe the atomic mirror formed by a chain of $N\gg 1$ identical two-level atoms strongly coupled to a one-dimensional waveguide and equally spaced by a distance $d_m$, as illustrated in Fig.\,\ref{fig:mirror}. For a monochromatic single photon impinging on the atomic mirror, the reflectance $R_m$ strongly depends on the lattice constant $d_m$ and the detuning $\Delta$ between the frequency of the input photon and $\omega_A$ \cite{Tsoi2008}. When the phase accumulated between two atoms of the chain $\omega_A d_m/v_g$ is a multiple of $\pi$, the reflectance for a photon close to resonance takes the form of a broadened Lorentzian~\cite{Chang2012}
\begin{equation}\label{eq:reflectance}
	R_m = \frac{1}{1+(\Delta/(N\gamma))^2} .
\end{equation}
The reflected frequency bandwidth is thus $N\gamma$, which is significantly enhanced by the large number of atoms in the chain, as shown in Fig.\,\ref{fig:reflectance}. Moreover, while the reflectance is sensitive to small fluctuations in the atomic positions $x_j+\delta x_j$ for some specific frequencies, the overall reflection coefficient of a broad bandwidth photon remains close to unity. This is an example of the well-known Bragg mirror \cite{Birkl1995,Schilke2011} and we will focus on this geometry for the rest of the paper. The presence of $N$ atoms in the mirror also leads to a superradiant decay at a much higher rate $N\gamma$, as recently observed experimentally \cite{Goban2015}. The time scale of the collective response of the atomic mirror is thus $1/(N\gamma)$, which will play a key role in the following.

We now consider the complete quantum cavity system formed by a central atom located between a set of two atomic mirrors~\cite{Chang2012}, as illustrated in Fig.\,\ref{fig:cavity}. {This central atom is identical to those of the mirrors}. An important parameter controlling the interaction between the central atom and the cavity is the phase shift acquired by a resonant photon across the cavity $\theta \equiv \omega_A d/v_g$. We will focus on the case $\theta = (2n+1)\pi$, where $n$ is an integer, such that $\omega_A$ corresponds to a natural mode of the cavity. In this configuration the central atom is located at an antinode of this mode, maximizing its interaction with the cavity electromagnetic field. All the atoms are assumed to be strongly coupled to the 1D waveguide and we first neglect losses due to the coupling with the environment. Effects of this noise will be discussed later.

The atomic transition frequency $\omega_A$ between ground $|g\rangle$ and excited $|e\rangle$ states is assumed to be much larger than the cutoff frequency of the waveguide. The dipole Hamiltonian describing the coupling between the atoms and the light field, under the rotating wave approximation, is then given by~\cite{Domokos2002}
\begin{eqnarray}\label{eq:Hamiltonian}
	\hat{H}_{dip} = - i\hbar \sum_{j=-N}^N\int_{0}^\infty\!d \omega\, g_\omega \Big[\hat{\sigma}^j_+ \Big(\hat{a}_\omega e^{i \omega x_j/v_g} +&& \hat{b}_\omega e^{-i \omega x_j/v_g} \Big)\nonumber\\&&-\mathrm{H.c.} \Big] ,
\end{eqnarray}
where $g_\omega$ is the coupling constant, $x_j$ is the position of atom $j$ with raising ladder operator $\hat{\sigma}^j_+\equiv |e\rangle_j\langle g|$, and $\hat{a}_\omega$ and $\hat{b}_\omega$ respectively annihilate right-going and left-going photons. In this work we use the Weisskopf-Wigner approximation and set $g_\omega=g_{\omega_A}$. Under this assumption, the decay rate of a single atom into each waveguide spatial mode is $\gamma\equiv 2\pi g_{\omega_A}^2$ \cite{Domokos2002}. It should be noticed that under typical experimental conditions, the cavity operates in a regime where $\gamma d/v_g \ll 1 \ll N$ (see Table \ref{Tab:tle}). In particular, in the limit $N\to\infty$ where the mirrors would ideally reflect all frequencies, one would expect to recover Rabi oscillation between the central atom and the cavity field as described in the framework of cavity quantum electrodynamics. However, as emphasized in the introduction, the \emph{key parameter} which determines the dynamics of the quantum cavity is given by the ratio between half the cavity round-trip time and the atomic mirrors response time $N\gamma d/v_g$.

\begin{table}
\caption{\label{Tab:tle} Experimental parameters for a trapped cesium atom \cite{Goban2015} and quantum dot \cite{Arcari2014} coupled to a photonic-crystal waveguide, and superconducting qubit coupled to a 1D coplanar waveguide transmission line \cite{Hoi2011}. For the distance between the mirrors we consider the range of values usually encountered in conventional CQED experiments with each type of TLEs.}
\begin{ruledtabular}
\begin{tabular}{ccccccc}
TLEs & $\omega_A$(GHz) & $2\gamma$(MHz) & $2\gamma/\gamma_0$\footnote{$\gamma_0$ is the decay rate into the environment, \textit{i.e.} outside of the waveguide modes.} & $v_g/c$ & $d$(mm) & $\gamma d/v_g$ \\
\hline
Cs atom & $2.1 \times 10^6$ & $32$ &$1.1$ &  $0.1$ & $1$ & $5.3 \times 10^{-4}$ \\
QD & $2\times 10^6$ & $6.2\times 10^3$ &  $63$ &$0.01$ & $10^{-2}$ & $1.0 \times 10^{-2}$ \\
SC & $7.1$ & $6 \times 10^2$ & $> 20$ & $0.5$ & $10$ &   $2 \times 10^{-2}$ \\
\end{tabular}
\end{ruledtabular}
\end{table}

\section{Time evolution}
We consider the situation when the atoms in the mirrors are initially in the ground state, while the central atom is prepared in the excited state.  The Schrödinger equation yields the following delay differential equations for the atomic excitation amplitudes (see Appendix \ref{app:A})
\begin{equation}\label{eomtheta}\begin{aligned} 
	\dot c_0(t) =& -\gamma c_0(t) - \gamma \sqrt{2 N} e^{i\theta/2} c_m(t-\tfrac d{2v_g})\Theta(t-\tfrac{d}{2v_g}) \\
	\dot c_m(t) =& -\gamma \sqrt{2 N} e^{i\theta/2} c_0(t-\tfrac{d}{2v_g})\Theta(t-\tfrac{d}{2v_g}) \\
	&- \gamma N \left[c_m(t) +e^{i\theta} c_m(t-\tfrac {d}{v_g})\Theta(t-\tfrac{d}{v_g})\right],  \end{aligned}
\end{equation}
where we have formally integrated the field variables, $c_m= \frac{1}{\sqrt{2N}}\sum_{j\neq0}(-1)^{(j+1)l} c_j$ depends on the excitation amplitude of the atoms forming the mirrors and $\Theta(t)$ is the Heaviside-step function. Here $l$ is the integer defined by $\omega_A d_m/ v_g=l\pi$. Moreover, the delay due the traveling time of a photon exchanged between the central atom and the two mirrors forming the cavity is included. 

These equations can be solved by Laplace transform. It is straightforward to show that the transform of $c_0(t)$ is
\begin{equation} \label{eq:c0s}\tilde{c}_0(s) = \frac{s + \gamma N(1+ e^{-s d/v_g+i\theta})}{(s+\gamma)(s+\gamma N)+ e^{-sd/v_g+i\theta}(s-\gamma)\gamma N}.
\end{equation} 
The excitation amplitude $c_0(t)$ is then obtained by taking the inverse Laplace transform. We show below that a compact analytical expression for $c_0(t)$ is possible in the regime of interest $\gamma d/v_g \ll 1\ll N$. As mentioned above we will mainly consider the case $\theta =(2n+1) \pi$, which is assumed in the rest of the text unless stated otherwise.

\subsection{Markovian regime}
We first apply the Markov approximation and neglect the delay $d/v_g$, which is valid if this delay is much smaller than the collective response time $1/(N \gamma)$ of the atomic mirrors. By setting $d=0$ in Eq.\,\eqref{eq:c0s} and obtaining the inverse Laplace transform, we find that, when $N \gg 1$, $c_0(t) = e^{-\gamma t/2}\cos(\sqrt{2N}\gamma t)$, which is consistent with the results of Ref.~\cite{Chang2012}. In other words, the central atom undergoes damped oscillations at the frequency $\Omega_{\text{Rabi}}\equiv\sqrt{2N} \gamma$, which however is not similar to the sustained Rabi oscillations often encountered in conventional CQED. Indeed, the associated decoherence rate $\gamma/2$ is found to be comparable to the decay rate of a single atom into the waveguide modes when the mirrors are not present, which suggests that the photon is actually not trapped in the cavity. Moreover, the oscillation frequency $\Omega_{\text{Rabi}}$ depends on the number of atoms in the mirrors, while the Rabi frequency usually depends on the modal volume of the field inside the cavity \cite{atphint1998}, which for our 1D architecture is given by the distance $d$ between the two mirrors. 

To provide further insight on why the oscillations in the Markovian regime do not show the usual features observed in conventional CQED, we recall that by neglecting the delay one has assumed that the central atom and the mirrors feel the influence of each other instantaneously, \textit{i.e.} $ d/v_g\ll 1/(\gamma N)$. 
Now for a propagating photon to be trapped in the cavity, its real-space distribution \cite{SFopt2005} must have a width at most comparable to $d$. {This implies} that the spread of this photon in momentum space is bounded by $\Delta \omega \gtrsim v_g/d$. On the other hand, $\Delta \omega$ has to be smaller than the reflection bandwidth $\gamma N$ of the atomic mirrors, which implies that $d/v_g \gtrsim 1/(\gamma N)$, where $1/(\gamma N)$ is the collective response time of the mirrors. This inequality manifestly contradicts the Markov approximation which assumes the delay to be smaller than any other dynamical time scale in the cavity.

The above discussion makes it clear that the dynamics of the system is inherently non-Markovian in the regime where the photon is trapped in the cavity. Moreover, the critical parameter that determines the transition between the two regimes is $N\gamma d/v_g$. 

\subsection{Non-Markovian regime}
\begin{figure}
\subfloat{
\includegraphics[width=0.25\textwidth]{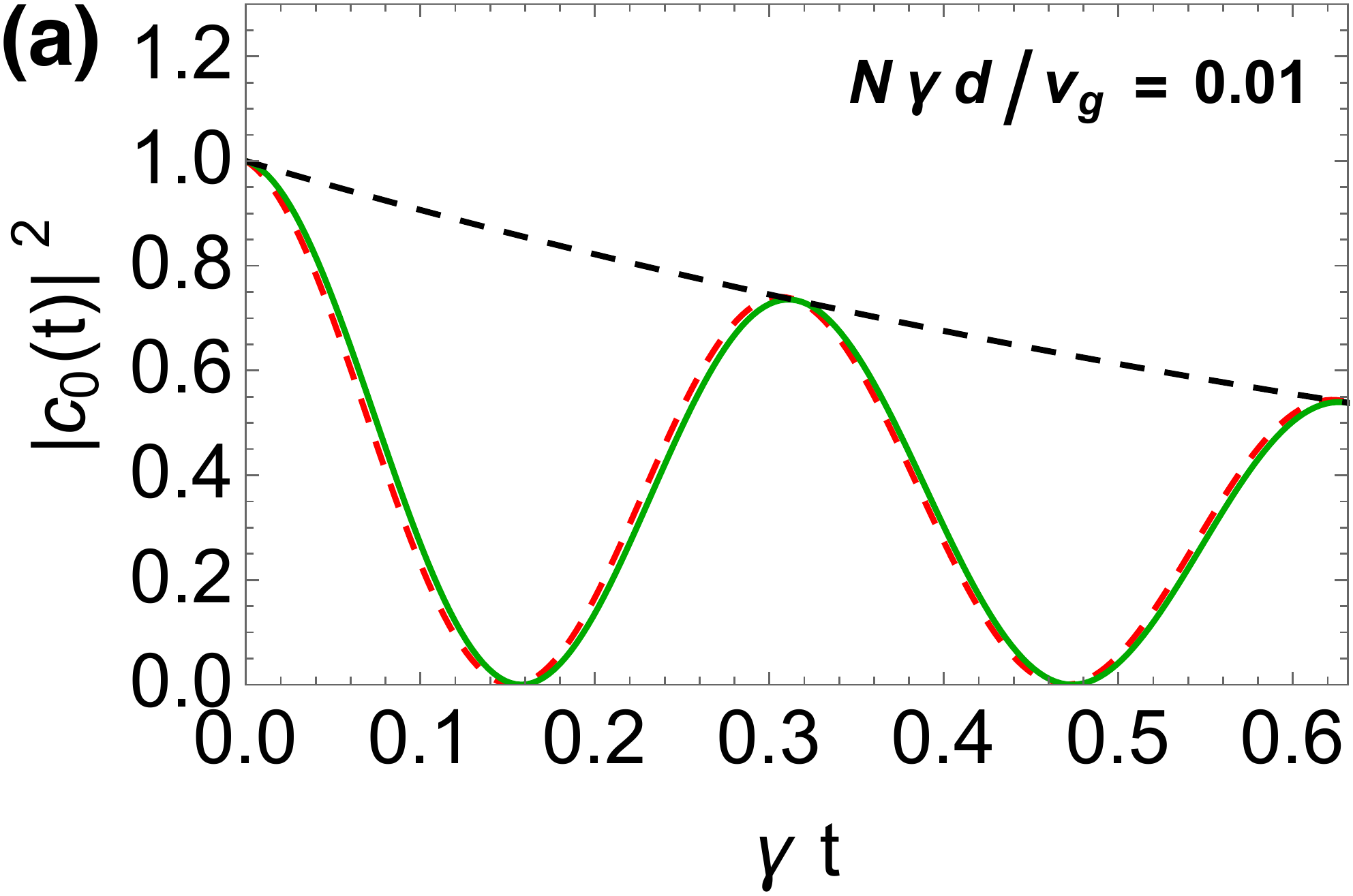}\label{fig:0.01}
}
\subfloat{
\includegraphics[width=0.25\textwidth]{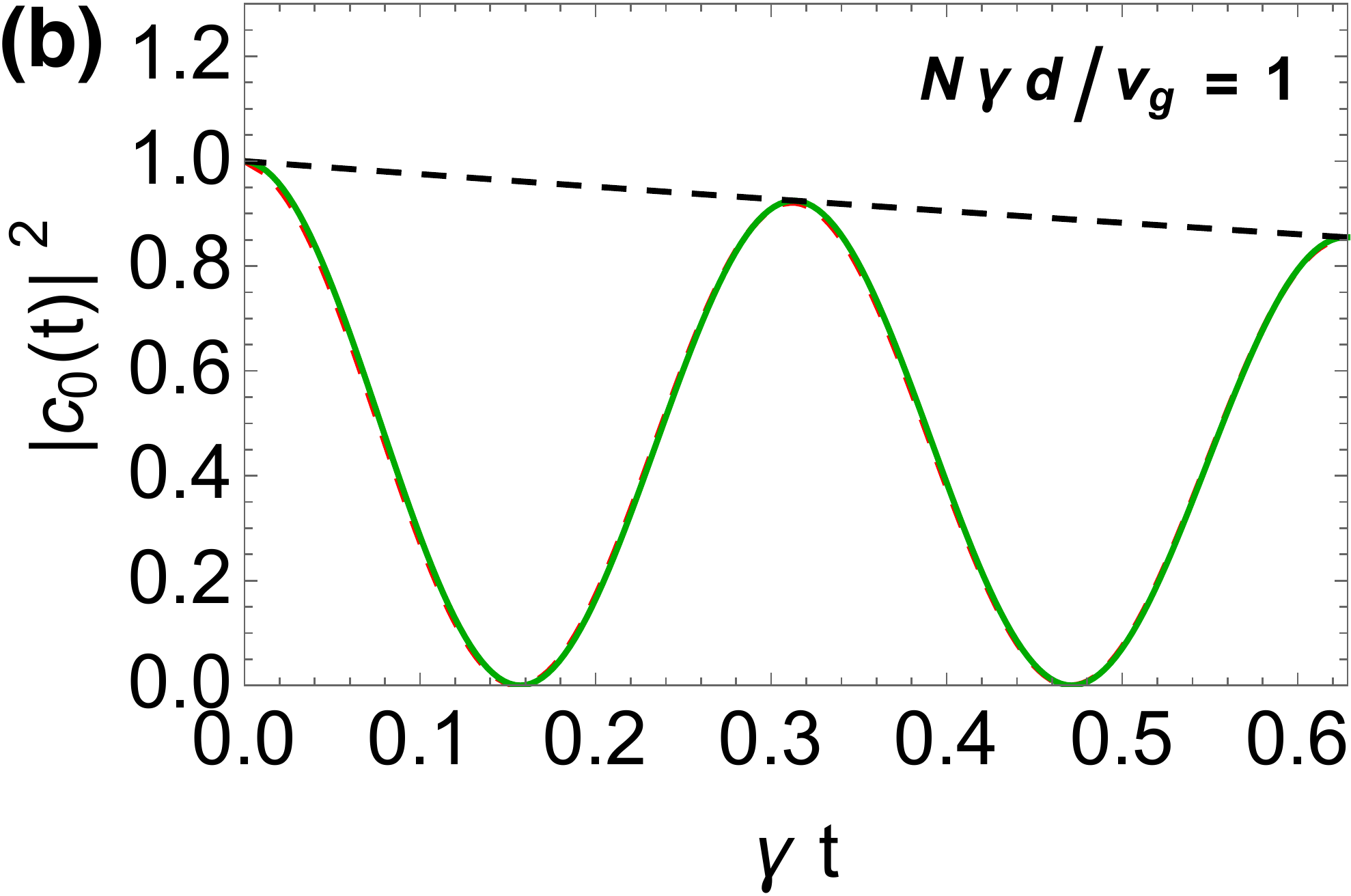}\label{fig:1}
}\\
\subfloat{
\includegraphics[width=0.25\textwidth]{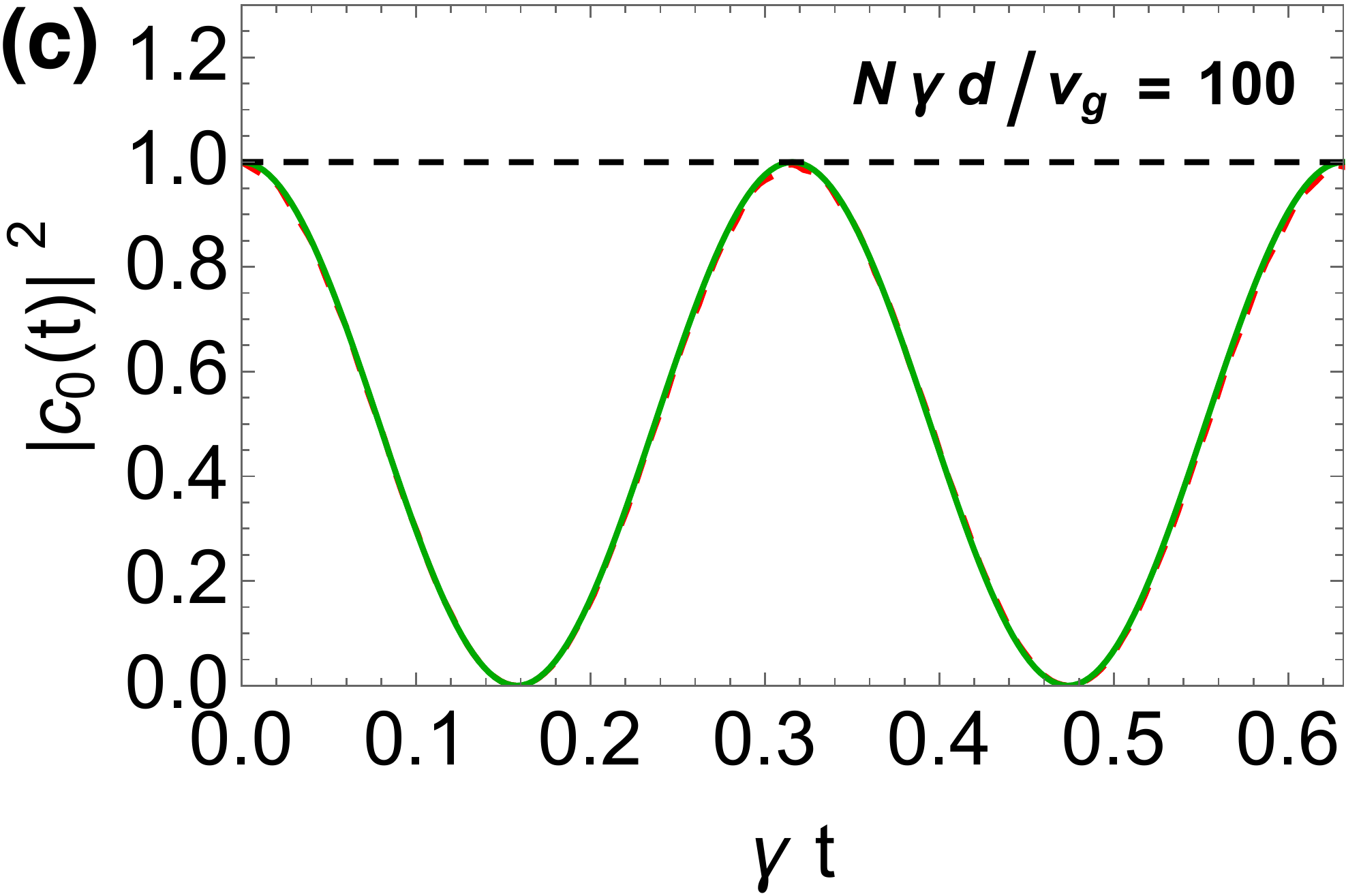}\label{fig:100}
}
\subfloat{
\includegraphics[width=0.25\textwidth]{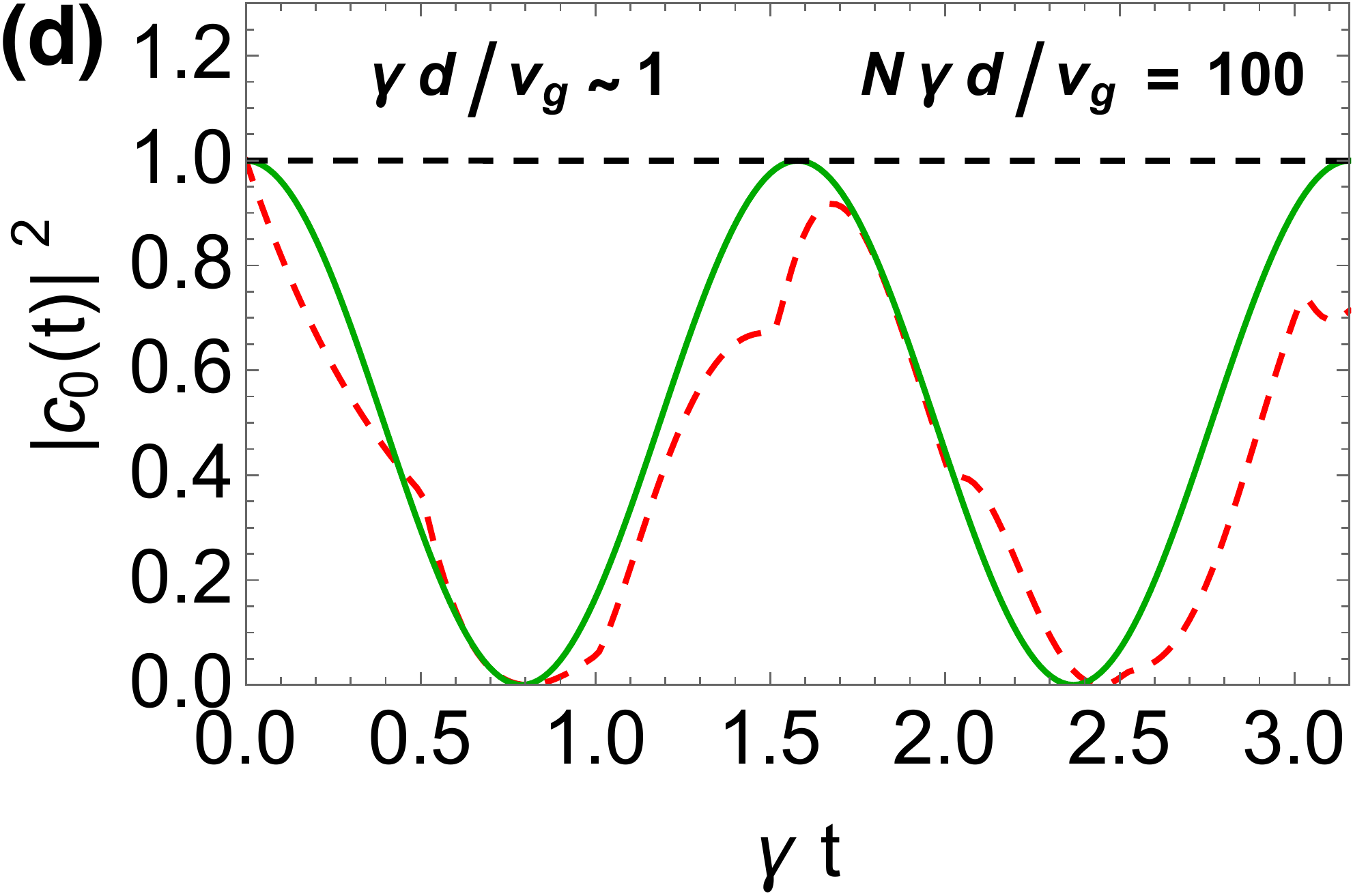}\label{fig:100bis}
}
\caption{\label{fig:oscillations}(color online).\quad Probability of excitation of the central atom for different values of the parameter $N\gamma d/v_g$. The green curve illustrates the analytical approximation \eqref{Eq:rabi} derived in the regime $\gamma d/v_g \ll 1\ll  N$, while the red dashed curve is obtained for any regime by directly inverting the Laplace transform in the form of \eqref{Eq:series}. The decaying envelope $e^{-\frac{\gamma t}{(1 + N\gamma d/v_g)^2 }}$ is plotted in dashed black for reference. \\(a) In the Markovian regime, we retrieve the damped oscillation obtained in \cite{Chang2012}, where the photon is not trapped in the cavity. $\gamma d/v_g=0.0002$  (b) In the transition regime, the lifetime of the photon in the cavity is significantly enhanced. $\gamma d/v_g=0.01$ (c) In the macroscopic regime, the photon is effectively trapped in the cavity and we observe sustained Rabi oscillation between the central atom and the cavity mode field. $\gamma d/v_g=0.02$ (d) For a large cavity length $\gamma d/v_g=0.5$, the retardation effects give rise to discontinuity at times $\gamma t=0.5 k$ with $k$ an integer. This is not captured {by Eq.\,\eqref{Eq:rabi}} which is limited to the regime $\gamma d/v_g \ll 1$.}
\end{figure}

We now study the time evolution of the central atom in the non-Markovian regime, where $N \gamma d/v_g$ may not necessarily be small. This requires solving Eq.\,\eqref{eomtheta} without neglecting the delay. The typical procedure for obtaining the inverse Laplace transform is to expand $\tilde{c}_0(s)$ in a geometric series and then taking the inverse transform of each term in the series \cite{methodsQO2002}. As shown in Appendix \ref{app:B}, this yields
\begin{equation}\label{Eq:series}
c_0(t)=\sum_{k=0}^{\infty} f_k(t-k d/v_g) \Theta(t-k d/v_g),
\end{equation}
where $f_k(t)$ are smooth functions in time, $f_0(0)=1$ and $f_{k\geq 1}(0)=0$. The exact form of $f_k(t)$ is not needed for our discussion and can be found in Appendix \ref{app:B}. The relevant observation here is that the Heaviside function gives rise to discontinuity in the slope of $c_0(t)$ at times $k d/v_g$, which correspond to the retardation times associated with successive exchanges of photons between the central atom and the mirrors (see Fig.\,\ref{fig:100bis}). This behavior is often encountered in the dynamic of quantum systems with delay~\cite{Milonni1974,Dorner2002}. 

However, under typical experimental circumstances, $\gamma d/v_g \ll 1$, and thus the evolution of the central atom appears smooth when the time scale of observation is comparable to $1/\gamma$. With this condition, it is also possible to obtain a compact analytical approximation for $c_0(t)$ which gives much more insight into the dynamics of the central atom than the exact series solution in Eq.~\eqref{Eq:series} does. In order to see this more clearly, we first discuss the ``macroscopic'' regime when $N\gamma d/v_g \gg 1$ so that we can set $N\rightarrow \infty$ in $\tilde{c}_0(s)$\footnote{Although $N\gg v_g/(\gamma d)$ in this regime, the assumption that the spatial extension of the mirror is small compared with the distance between the mirrors is still valid as long as $N d_m\ll d$. We can thus safely neglect any increase in the cavity mode volume due to penetration of the field into the mirrors \cite{Kimble2001}.}. As a result, the Laplace transform in Eq.~\eqref{eq:c0s} is simplified to 
\begin{equation} \label{eq:c0sainf} \tilde{c}_0(s) = \left(s+\gamma\frac{1+e^{-sd/v_g}}{1-e^{-sd/v_g}}\right)^{-1}. 
\end{equation}

Instead of geometric series expansion, another approach for obtaining the inverse Laplace transform is to identify the poles $s_j$ of $\tilde{c}_0(s)$. The excitation amplitude is then readily obtained via Cauchy's theorem as $c_0(t)=\sum_j r_j$ where $r_j$ is the residue of $\tilde{c}_0(s) e^{st}$ evaluated at the pole $s_j$. We find that the poles of $\tilde{c}_0(s)$ are simple poles confined to the imaginary axis, \textit{i.e.} $s_j=iy_j$ where $y_j$ is a real number. Moreover, the exact location of the poles are given by the solutions of the equation $y = \gamma \cot(\frac{y d}{2v_g})$, which leads to an infinitude of poles that have reflection symmetry through the coordinate origin. The residue at the pole $s_j$ is $r_j\approx e^{s_j t}/[1-s_j^2 d/(2\gamma v_g)]$. When $\gamma d/v_g\ll 1$, the two poles nearest to the origin are $s_{\pm1}\approx \pm i\sqrt{2\gamma v_g/d}$, and the contribution of these two poles to $c_0(t)$ is approximately $\cos\left(\sqrt{2\gamma v_g/d} \ t\right)$. The contribution of the other poles are shown to be bounded above by $\gamma d/(6v_g)$, which is negligible (see Appendix \ref{app:B}). Therefore, we obtain for the central atom excitation amplitude
\begin{equation}
c_0(t)\approx \cos\left(\sqrt{\frac{2\gamma v_g}{d}}t\right) .
\end{equation}
This results implies that in the macroscopic regime $N\gamma d/v_g \gg 1$, the central atom undergoes sustained Rabi oscillation without decoherence. This is understandable since the atomic mirrors have a reflection bandwidth that is much larger than the frequency width of the photon trapped in the cavity. {Moreover}, the Rabi oscillation is proportional to $1/\sqrt{d}$, which is expected since in one dimension $d$ plays the role of the modal volume of the field in the cavity.

Now we consider the more general situation when $N\gamma d/v_g$ may have arbitrary value. The two poles nearest to the coordinate origin are shifted to 
\begin{equation}
s_{\pm 1}\approx -\gamma/(2(1+N\gamma d/v_g)^2)\pm i\gamma \sqrt{2N/(1+N\gamma d/v_g)},
\end{equation}
and their contribution yields
\begin{equation} \label{Eq:rabi}
c_0(t) \approx e^{-\frac{\gamma t}{2(1 + N\gamma d/v_g)^2 }}\cos\left(\sqrt{\frac{2 N}{1+N\gamma d/v_g} } \gamma t\right). 
\end{equation}
The contribution of the other poles are again negligible in the regime of interest $\gamma d/v_g \ll 1\ll N$. Both the Markovian limit, $N\gamma d/v_g \ll 1$, and the macroscopic limit, $N\gamma d/v_g \gg 1$, are correctly retrieved from Eq.\,\eqref{Eq:rabi}. Fig.\,\ref{fig:oscillations} shows a comparison between our compact analytical approximation obtained in Eq.\,\eqref{Eq:rabi} and the exact form of $c_0(t)$ in Eq.\,\eqref{Eq:series} for various values of $\gamma d/v_g$ and $N$. The approximation is indeed very accurate in the regime of small $\gamma d/v_g$.

From Eq.\,\eqref{Eq:rabi}, one sees that the cavity loss rate is given by $\kappa = \gamma/{(1 + N\gamma d/v_g)^2}$ while the Rabi frequency reads $\Omega_{\text{Rabi}}=\gamma \sqrt{\frac{2 N}{1+N\gamma d/v_g} }$. We define the \textit{critical size of the mirror} $N_c=v_g/(\gamma d)$ that signals the transition from the delocalization to the localization of the photon in the cavity. The cavity loss rate is greatly suppressed when $N\gtrsim N_c$. One can refer from Table \ref{Tab:tle} that $N_c\approx 1900$ for the first configuration with Cesium atoms, $N_c\approx 100$ for quantum dots and {$N_c\approx 50$} for superconducting qubits. While the first configuration seems to require a large number of atoms, it should be noted that the critical size $N_c$ can be reduced by an order of magnitude by slowing light in the photonic-crystal waveguide \cite{Baba2008,Li2012} or increasing the coupling strength between the atoms and the waveguide. Besides, as discussed previously, small position fluctuations do not change much the average reflection of the mirrors forming the cavity; therefore, one does not need a significant extra number of atoms in order to trap the photon in the cavity.

\section{Effects of detuning and loss to environment}
We have studied the Rabi oscillation of the central atom inside a quantum cavity when the atom is on resonance with one of the natural modes of the cavity, that is, $\omega_A=(2n+1) \pi v_g/d$. We now study the effect of a small detuning between the atomic transition frequency and the cavity modes and show that our model recovers the behaviour predicted by CQED with classical mirrors. In particular, the phase accumulated across the cavity by a photon on resonance with the atom now reads $\theta = (2n+1)\pi + \phi$ with $\phi \ll \pi$. The frequency of the main cavity mode interacting with the trapped atom is  $\omega_c \equiv (2n+1)\pi v_g/d$, yielding a detuning $\Delta \equiv \omega_A -\omega_c = \phi v_g/d$. Focusing on the macroscopic regime (see Appendix \ref{app:C} for the case of arbitrary $N\gamma d/v_g$), the poles of $\tilde{c}_0(s)$ in Eq.\,\eqref{eq:c0s} are now given by the solutions of the equation $y = \gamma \cot(\frac{(y-\Delta)d}{2v_g})$. The two main poles are thus shifted to $s_{\pm1} =  i\Delta /2 \pm i \sqrt{\Omega_0^2 + (\Delta /2)^2}$, where $\Omega_0$ is the Rabi frequency obtained previously without detuning, and their contribution yields
\begin{equation} 
	c_0(t) \approx e^{i\frac\Delta2t}\Big(\cos(\Omega t) - i \sin(\Omega t)\frac\Delta{2\Omega}\Big) 
\end{equation}
where the generalized Rabi frequency reads $\Omega = \sqrt{\Omega_0^2 + (\Delta /2)^2}$, similarly to conventional CQED setups \cite{Agarwal2012}.

In realistic circumstances, each atom coupled to the 1D waveguide also emits independently into the free-space environment with rate $\gamma_0$. In practice, this induces additional loss in the cavity and impacts the decoherence rate of the Rabi oscillation obtained in Eq.\,\eqref{Eq:rabi} which would now read $\kappa +\gamma_0$. We thus see that the decoherence rate is now bounded from below by the loss to the environment at $\gamma_0$, no matter how the key parameter $N\gamma d/v_g$ is optimized. However, waveguide-QED setups allow the realization of remarkably high coupling efficiency between the atoms and the waveguide modes, which implies that the period of Rabi oscillation is much smaller than the time scale of the loss for large $N$. Indeed, for the three configurations in Table \ref{Tab:tle}, the ratio between $\Omega_{\text{Rabi}}$ and the total decoherence rate is around $21$, $35$ and $20$ at the critical number of atoms $N_c$ in the mirrors . Hence, one expects to observe many cycles of Rabi oscillation before the photon eventually leaks to the environment.  The cooperativity is given by $\eta\equiv \Omega^2_\text{Rabi}/(\gamma_0 \kappa)=2N(1 + N\gamma d/c)\gamma/\gamma_0$, which is very large when $N$ is close to $N_c$. Large cooperativity is essential for many quantum information processing applications based on strong atom-photon coupling such as the nanophotonic quantum phase switch \cite{Tiecke2014} and the atom-photon quantum logic gate \cite{Reiserer2014}.

\section{Conclusion}
When comparing this work with the description of a cavity made of classical mirrors, the reader might be surprised by the apparent need of a non-Markovian theory in order to reproduce standard Rabi oscillation. However, one should notice that the non-Markovianity is required to build up the cavity standing wave out of propagating modes. This is not in contradiction with standard CQED where the standing wave structure is assumed \textit{a priori} by imposing that the electric field vanishes at the mirrors position. In the latter scenario, one then assumes that the evolution occurs simultaneously throughout the cavity, with the photons being delocalized in the standing-wave spatial mode. Notably, this description is valid in the regime $\gamma d/v_g\ll 1$ where retardation effects are negligible \cite{Parker1987}. A non-Markovian treatment is required only when we also consider the dynamics of the mirrors whose characteristic time scale is much smaller than the delay, as can be seen in our calculation.

In this work, we have derived the Rabi frequency and the decoherence rate of the Rabi oscillation undergone by an atom placed inside a quantum cavity formed by atomic mirrors. We found a simple condition for testing whether the dynamics of the system can be correctly described with the Markov approximation. Our approach is valid for both the Markovian regime, where the photon's travel time between the mirrors is neglected, and the non-Markovian regime, where we found that the photon can be trapped inside the cavity. In the latter regime, we have found that sustained Rabi oscillation, as achieved in conventional CQED with high finesse mirrors, can be observed and the experimental parameters required for achieving it are estimated.

\begin{acknowledgements}
We thank Jason Twamley for insightful discussions and Marc-André Dupertuis for helpful comments. This research is supported by the National Research Foundation (partly through its Competitive Research Programme, Award No. NRF-CRPX-20YY-Z) and the Ministry of Education, Singapore.
\end{acknowledgements}

\newpage
\onecolumngrid
\begin{appendix}

\section{Deriving the coupled delay-differential equations}\label{app:A}
As illustrated in Fig.~\ref{fig:cavity}, our system is constituted of two chains of $N$ atoms separated by a distance $d$, with a spacing of $d_m$ between neighboring atoms in each chain, and an additional central atom located between both chains. We consider the situation when the length of the chain is much smaller than the distance between the two atomic mirrors. The Hamiltonian in the interaction picture reads 
\begin{equation} \hat{H}_{dip} = - i\hbar \sum_{j=-N}^N\int_{0}^\infty\!d \omega\, g_\omega \Big[\hat{\sigma}^j_+ \Big(\hat{a}_\omega e^{i \omega x_j/v_g} + \hat{b}_\omega e^{-i \omega x_j/v_g} \Big)e^{-i (\omega-\omega_A)t} -\mathrm{H.c.} \Big] ,
\end{equation} 
where $x_j$ is the position of the $j$th atom. The state of the system can be written as 
\begin{align}
\ket{\Psi(t)} = &\sum_{j=-N}^{N}c_j(t)\hat{\sigma}^j_+\ket \varnothing +\int_{0}^\infty\!d \omega\,(c_a(\omega,t)\hat{a}^\dagger_\omega + c_b(\omega,t)\hat{b}^\dagger_\omega)\ket\varnothing,
\end{align}
where $\ket \varnothing$ indicates the combined atom-field vacuum, which is the state when there is no photon in the waveguide and all the atoms are in the ground state. The initial state where only the central atom is in the excited state corresponds to $c_j(0)=\delta_{0,j}$ and $c_a(\omega,0)=c_b(\omega,0)=0$. The Schrödinger equation then leads to the following set of equations 
\begin{align}
\label{eqa}\dot c_a(\omega,t) &= \sum_{j=-N}^{N}c_j(t) g_\omega e^{i (\omega - \omega_A)t}e^{-i\omega x_j/v_g}, \\\label{eqb}
 \dot c_b(\omega,t) &= \sum_{j=-N}^{N}c_j(t) g_\omega e^{i (\omega - \omega_A)t}e^{i\omega x_j/v_g}, \\\label{eqc} 
 \dot c_j(t) &= - \int_{0}^\infty\!d \omega\, g_\omega e^{-i (\omega - \omega_A)t}\left[c_a(\omega,t)e^{i\omega x_j/v_g} + c_b(\omega,t)e^{-i\omega x_j/v_g}\right].  
\end{align}
In the Weisskopf-Wigner approximation one can replace $g_\omega$ by $g_{\omega_A}$, which is done in the rest of the text. Integrating formally the differential equations for $c_a(\omega,t)$ and $c_b(\omega,t)$ and then inserting them into the equations for $c_j(t)$, we obtain the following delay-differential equations for the atomic excitation amplitudes
\begin{equation}\dot c_j(t) = -\gamma \sum_{j'=-N}^N e^{i \omega_A |x_j - x_{j'}|/v_g} c_{j'}(t - |x_j-x_{j'}|/v_g).\end{equation}
For the configuration of our system, the atoms are located at the following positions
\[
\begin{cases}
x_0 &=0, \\
x_j &= (j+1)d_m - d/2 \quad \text{for} \ -N\leq j \leq -1,  \\
x_j &=(j-1) d_m + d/2 \quad \text{for} \ \ \ \ \ \ 1\leq j \leq N.
\end{cases}
\]

Since the spatial extension of each atomic mirror is much smaller than the distance between the two mirrors, we can neglect the time delay due to the distance between any two atoms in the \textit{same mirror}. Moreover, we focus on the configurations where $\omega_A d_m/v_g = l \pi$ for two adjacent atoms in the same mirror, where $l$ is an integer.  In this scenario one can check that every coefficient $(-1)^{(j+1)l} c_j(t)$ for $j\neq 0$ are equal. Denote the excitation amplitude of the cavity by $c_m(t) = \frac{1}{\sqrt{2N}}\sum_{j\neq0}c_j(t)(-1)^{(j+1)l}$, we then obtain the two coupled delay-differential equations for the central atom and the cavity
\begin{eqnarray}
\dot c_0(t) &=& -\gamma c_0(t) - \gamma \sqrt{2N} e^{i \theta/2}c_m(t - \tfrac{d}{2v_g})\Theta(t-\tfrac{d}{2v_g}), \\
\dot c_m(t) &=&- \gamma \sqrt{2N}e^{i \theta/2} c_0(t-\tfrac{d}{2v_g})\Theta(t-\tfrac{d}{2v_g}) - \gamma N \left[c_m(t) + e^{i \theta}  c_m(t-\tfrac{d}{v_g})\Theta(t-\tfrac{d}{v_g})\right]. \end{eqnarray}
For $\theta = (2n+1)\pi$, the Laplace transform of $c_0(t)$, denoted by $\tilde{c}_0(s)$, is 
\begin{equation}\label{laplace} \tilde{c}_0(s) = \frac{s+\gamma N(1-e^{-sd/v_g})}{(s+\gamma)(s+\gamma N)}\frac{1}{1-e^{-sd/v_g}\frac{(s-\gamma)\gamma N}{(s+\gamma)(s+\gamma N)}}, \end{equation}
and $c_0(t)$ can be obtained by inverting the Laplace transform, which in our case can be done by summing the residues at the poles of $e^{st} \tilde{c}_0(s)$.

\section{Solving the equation}\label{app:B}

Let us expand the second term in the expression of $\tilde{c}_0(s)$ in a geometric series 
\begin{align*}\frac{1}{1-e^{-sd/v_g}\frac{(s-\gamma)\gamma N}{(s+\gamma)(s+\gamma N)}} = \sum_{k=0}^\infty e^{-s kd/v_g}\left[\frac{(s-\gamma)\gamma N}{(s+\gamma)(s+\gamma N)}\right]^k. \end{align*} 
We then get 
\begin{eqnarray*} \tilde{c}_0(s) &=& \frac{s+\gamma N}{(s+\gamma)(s+\gamma N)} \sum_{k=0}^\infty e^{-s kd/v_g}\left[\frac{(s-\gamma)\gamma N}{(s+\gamma)(s+\gamma N)}\right]^k \\ &-&\frac{\gamma N}{(s+\gamma)(s+\gamma N)} \sum_{k=0}^\infty e^{-s (k+1)d/v_g}\left[\frac{(s-\gamma)\gamma N}{(s+\gamma)(s+\gamma N)}\right]^k,
\end{eqnarray*}
and by taking the inverse Laplace transform term by term we have 
\begin{equation} \label{eq:seriesSol}
c_0(t) = \sum_{k=0}^\infty f_k(t-k d/v_g) \Theta(t-k d/v_g) 
\end{equation}
with 
\begin{eqnarray*}f_0 &=&  \mathcal L^{-1}\Big\{\frac{1}{s+\gamma}\Big\} = e^{-\gamma t}, \\
f_{k\geq1} &=& \mathcal L^{-1}\Bigg\{\frac{1}{s+\gamma}\left[\frac{(s-\gamma)\gamma N}{(s+\gamma)(s+\gamma N)}\right]^k -\frac{\gamma N}{(s+\gamma)(s+\gamma N)}\left[\frac{(s-\gamma)\gamma N}{(s+\gamma)(s+\gamma N)}\right]^{k-1} \Bigg\}
\\ &=&\mathcal L^{-1}\Big\{ \Bigg( \frac1{s+\gamma} - \frac{1}{ s-\gamma}\Bigg)\left[\frac{(s-\gamma)\gamma N}{(s+\gamma)(s+\gamma N)}\right]^k \Bigg\}, \end{eqnarray*}
where $\mathcal L^{-1}$ denotes the inverse Laplace transform and we made use of $\mathcal L^{-1}\{e^{-skd/v_g}F\}(t) = \mathcal L^{-1}\{F\}(t-kd/v_g)\Theta(t-kd/v_g)$. One can check that $f_k$ is a smooth function of time since its Laplace transform involves only polynomials, and furthermore $f_k(0)=0$ for $k\geq 1$. The evolution for $t<d/v_g$ is that of a single atom in the waveguide, which is due to causality: For $t<d/(2v_g)$, the photon wavepacket emitted by the central atom has not yet reached the mirrors; for $t \geq d/(2v_g)$, the mirrors can be excited by the influence of the central atom, however their back-action on the central atom will be delayed by another duration of $d/(2v_g)$.

The series solution for $c_0(t)$ given in \eqref{eq:seriesSol} is exact but does not give much insight into the dynamics of the central atom. We now derive the compact analytical approximation of $c_0(t)$ in the regime of interest $\gamma d/v_g \ll 1 \ll N$. First, in the Markovian regime when $N \gamma d/v_g \ll 1$, the solution is straightforwardly obtained by setting $d/v_g=0$ in Eq.~\eqref{laplace} and taking the inverse transform, which yields 
\begin{equation} 
c_0(t) = e^{-\frac{\gamma t}2}\cos\left( \sqrt{2N}\gamma t\right) + \mathcal{O}\left(\frac{1}{N}\right).
\end{equation}

\begin{figure}
\begin{center}
\includegraphics[width=9cm]{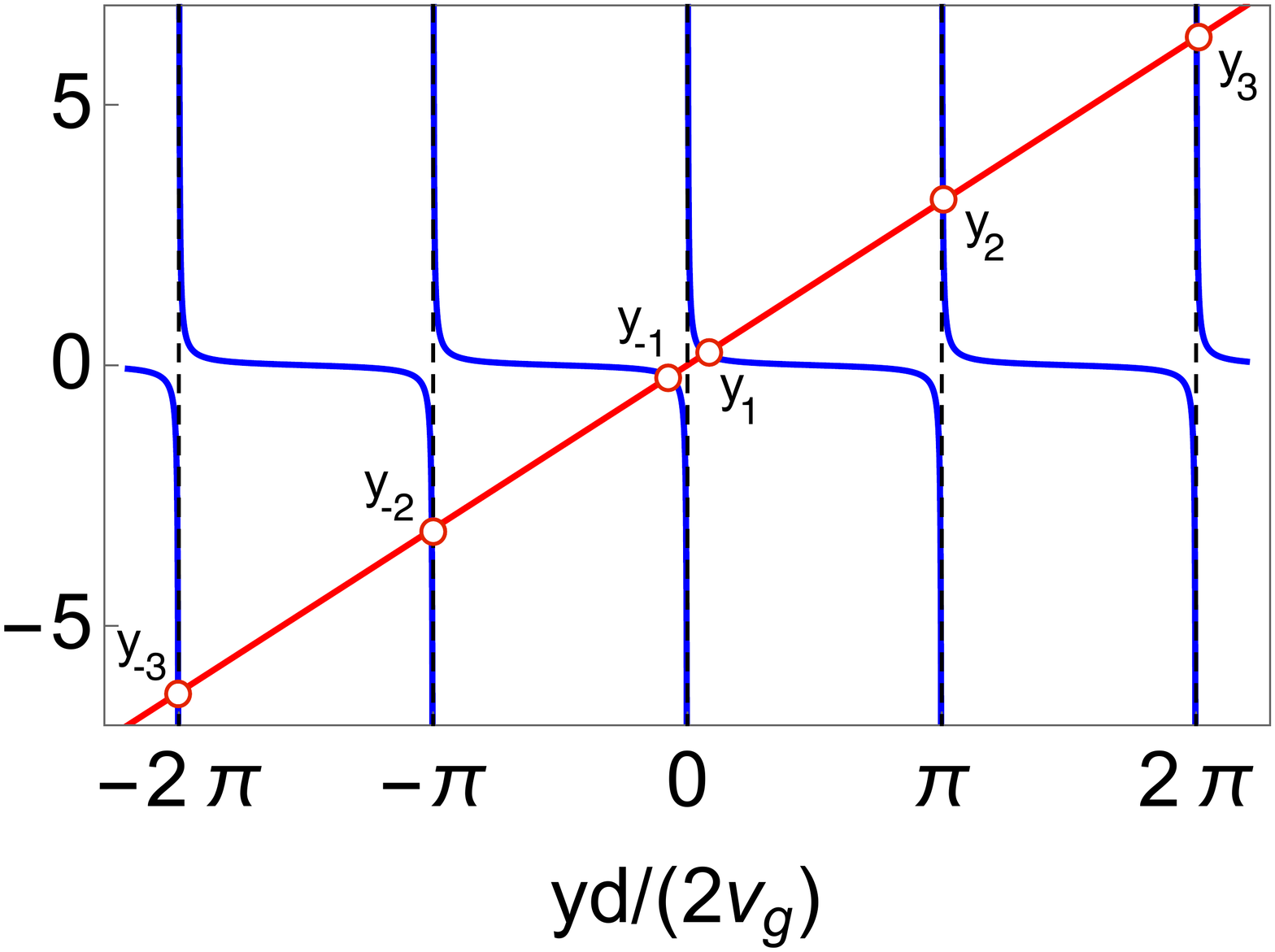}
\caption{\label{solpoles}(color online). $y^\prime$ in red and $\cot(y^\prime)\gamma d/(2v_g)$ in blue as a function of the dimensionless variable $y^\prime=y d/(2 v_g)$, for $\gamma d/v_g=0.1$. The intersections correspond to the poles of $\tilde{c}_0(s)$.}
\end{center}
\end{figure}

In the opposite extreme, the ``macroscopic" regime when $N \gamma d/v_g\gg1$, we set $N\rightarrow \infty$ in Eq.~\eqref{laplace} and obtain
\begin{equation}\label{eqmacro}   \tilde{c}_0(s) = \frac{1-e^{-sd/v_g}}{s+\gamma-(s-\gamma)e^{-sd/v_g}}. 
\end{equation} 
The poles, which are given by the zeros of the denominator, must obey $\frac{s+\gamma}{s-\gamma} = e^{-sd/v_g}$, and thus $\frac{|s+\gamma|}{|s-\gamma|} = e^{-\mathrm{Re}(s) d/v_g}$. With a bit of algebra one can show that this constraint is satisfied if and only if $\mathrm{Re}(s) = 0$. Hence the poles of $\tilde{c}_0(s)$ must lie on the imaginary axis. Denote $y=\mathrm{Im}(s)$, we have
\begin{equation}
\tilde{c}_0(i y) = \frac{-i}{y-\gamma \cot(\frac{yd}{2v_g})},
\end{equation} 
and the poles are given by the solution of $y = \gamma \cot(\frac{y d}{2v_g})$. Note  that both functions are odd, so if $y$ is a solution then $-y$ is also a solution (see figure \ref{solpoles}). In the limit $\gamma d/v_g \ll 1$, the two solutions closest to the origin are given by $y_{\pm 1} \approx \pm \sqrt{\frac{2\gamma v_g}{d}}$. There are also other solutions close to each singularity of $ \cot(\frac{y d}{2v_g})$, denoted by $y_j, j=\pm 2,\pm 3, \dots$. From Fig.~\ref{solpoles} we see that $|y_j| > (|j|-1)2 \pi v_g/d$. 

We can now obtain $c_0(t)$ by evaluating the residue at each pole 
\begin{eqnarray}\label{eq:residuesLim} c_0(t) &=& \sum_{j}\lim_{y\rightarrow y_j} \frac{y-y_j}{y-\gamma\cot(\frac{yd}{2v_g})}e^{iyt}\nonumber\\
&=& \sum_{j}\frac{e^{iy_j t}}{1+\frac{\gamma d}{2v_g}+\frac{\gamma d}{2v_g}\cot(\frac{y_j d}{2v_g})^2}\nonumber\\
&=& \frac{e^{iy_{+1}t}+e^{iy_{-1}t}}{1+\frac{\gamma d}{2v_g} + y_{\pm 1}^2\frac{d}{2\gamma v_g}} + \sum_{j=2}^\infty\frac{e^{iy_j t}+e^{iy_{-j}t}}{1+\frac{\gamma d}{2v_g} + y_j^2\frac{d}{2\gamma v_g}}. \end{eqnarray}
The sum can be bounded  as follows
\begin{equation}\label{eq:residuesBorn} \Bigg|\sum_{j=2}^\infty \frac{2\cos(y_j t)}{1+\frac{\gamma d}{2v_g} + y_j^2\frac{d}{2\gamma v_g}} \Bigg| < \sum_{j=2}^\infty\frac{2}{y_j^2\frac{d}{2\gamma v_g}} < \frac{\gamma d}{v_g \pi^2}\sum_{j=2}^\infty\frac{1}{(j-1)^2} = \frac{1}{6}\frac{\gamma d}{v_g}
\end{equation}
since $|y_j| > (|j|-1)2 \pi v_g/d$ for $j\neq \pm 1$. Therefore, we arrive at the analytical approximation
\begin{equation} c_0(t) = \cos\left(\sqrt{\frac{2\gamma v_g}{d}}t\right)+\mathcal O\left(\frac{\gamma d}{v_g}\right).
\end{equation}
One observes that the main contribution to $c_0(t)$ is from the two poles nearest to the coordinate origin. We find that there is a simpler procedure to obtain the position of these two poles approximately. This is done by expanding $e^{-sd/v_g}\approx 1-s d/v_g+(s d/v_g)^2/2$ in the expression of $\tilde{c}_0(s)$ in Eq.~\eqref{eqmacro} and finding the zeros of the resulting denominator (which is a cubic polynomial but only two of its zeros correspond to the two main poles we are interested in). 

Now we study the general case when $N \gamma d/v_g=a$ where $a$ can have any finite value. First replace $N=a v_g/(d\gamma)$ in the expression of $\tilde{c}_0(s)$ in Eq.~\eqref{laplace}, and then expand $e^{-sd/v_g}$ to second order. Solving for the zeros of the resulting denominator, we find that the two main poles are now shifted to
\begin{equation}
s_{\pm 1}\approx -\gamma/[2(1+a)^2]\pm i \gamma \sqrt{\frac{2 a v_g/(d\gamma)}{1+a}},
\end{equation}
and their contribution yields
\begin{equation}
c_0(t) = e^{-\frac{\gamma t}{2(1 + a)^2 }}\cos\left(\sqrt{\frac{2 a v_g/(d\gamma)}{1+a} } \gamma t\right) +\mathcal O\left(\frac{\gamma d}{v_g}\right),
\end{equation}
which is Eq.~\eqref{Eq:rabi}.

\section{Effects of Detuning}\label{app:C}

Let us consider a detuning $\Delta$ between the cavity and the atomic frequency. Then $\theta = (2n+1)\pi + \phi$, where $\phi = \Delta d/v_g$. The Laplace transform now takes the form \begin{equation}\label{laplacedetuning} \tilde{c}_0(s) = \frac{s+\gamma N(1-e^{i \Delta d/v_g}e^{-sd/v_g})}{(s+\gamma)(s+\gamma N)}\frac{1}{1-e^{i \Delta d/v_g}e^{-sd/v_g}\frac{(s-\gamma)\gamma N}{(s+\gamma)(s+\gamma N)}}. \end{equation}

We first focus on the good cavity limit when the photon is trapped, which happens when $N\gamma d/v_g \gg 1$. After setting $N\rightarrow \infty$ we have \begin{equation}\label{eqmacro2}   \tilde{c}_0(s) = \frac{1-e^{i\Delta d/v_g}e^{-sd/v_g}}{s+\gamma-(s-\gamma)e^{i\Delta d/v_g}e^{-sd/v_g}}. 
\end{equation} 
As in the case without detuning, the poles lay on the imaginary axis and satisfy $y = \gamma \cot(\frac{(y-\Delta)d}{2v_g})$. Hence, for $\Delta d/v_g \ll \pi$, the two main poles are now given by
\begin{equation}
	s_{\pm 1} = i\Delta/2 \pm i\sqrt{\Omega_0^2 + (\Delta/2)^2}
\end{equation}
with $\Omega_0 = \sqrt{2\gamma v_g/d}$. Repeating the derivation of \eqref{eq:residuesLim}, the contribution of the other poles is still of order $\mathcal O(\gamma d/v_g)$ and the inverse Laplace transform reads
\begin{equation} c_0(t) = e^{i\frac\Delta2 t}\Big(\cos(\Omega t) - i \sin(\Omega t) \frac{\Delta}{2\Omega}\Big)+\mathcal O(\gamma d/v_g)
\end{equation}
which is consistent with cavity-QED. The probability of excitation is then
\begin{equation}
	P_0(t)= \frac{\cos(\Omega t) + (\frac\Delta{2\Omega_0})^2}{1+(\frac\Delta{2\Omega_0})^2}. \end{equation}

\begin{figure}[t]
\subfloat{
\includegraphics[width=0.3\textwidth]{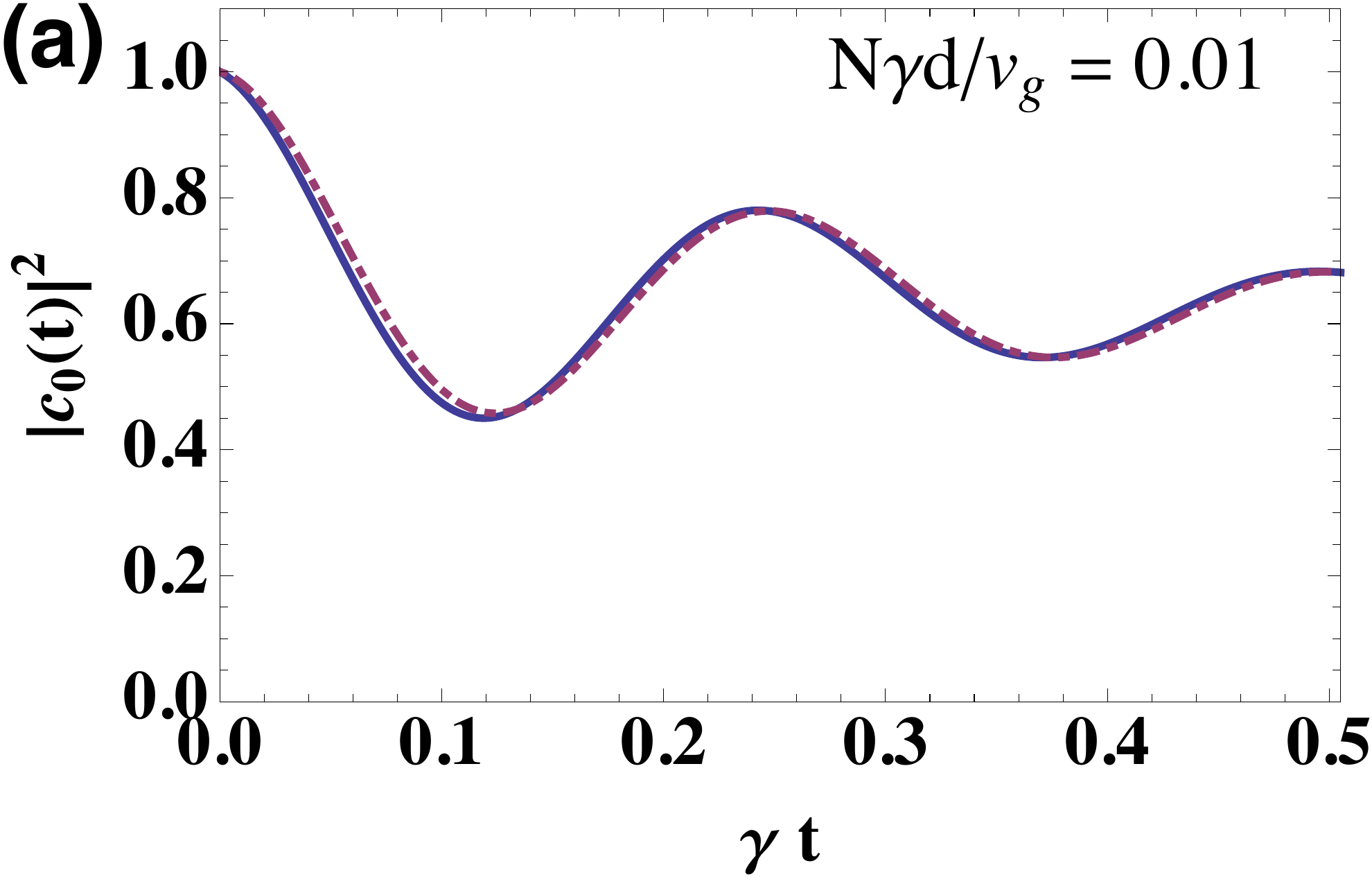}\label{fig:App0.01}
} \subfloat{
\includegraphics[width=0.3\textwidth]{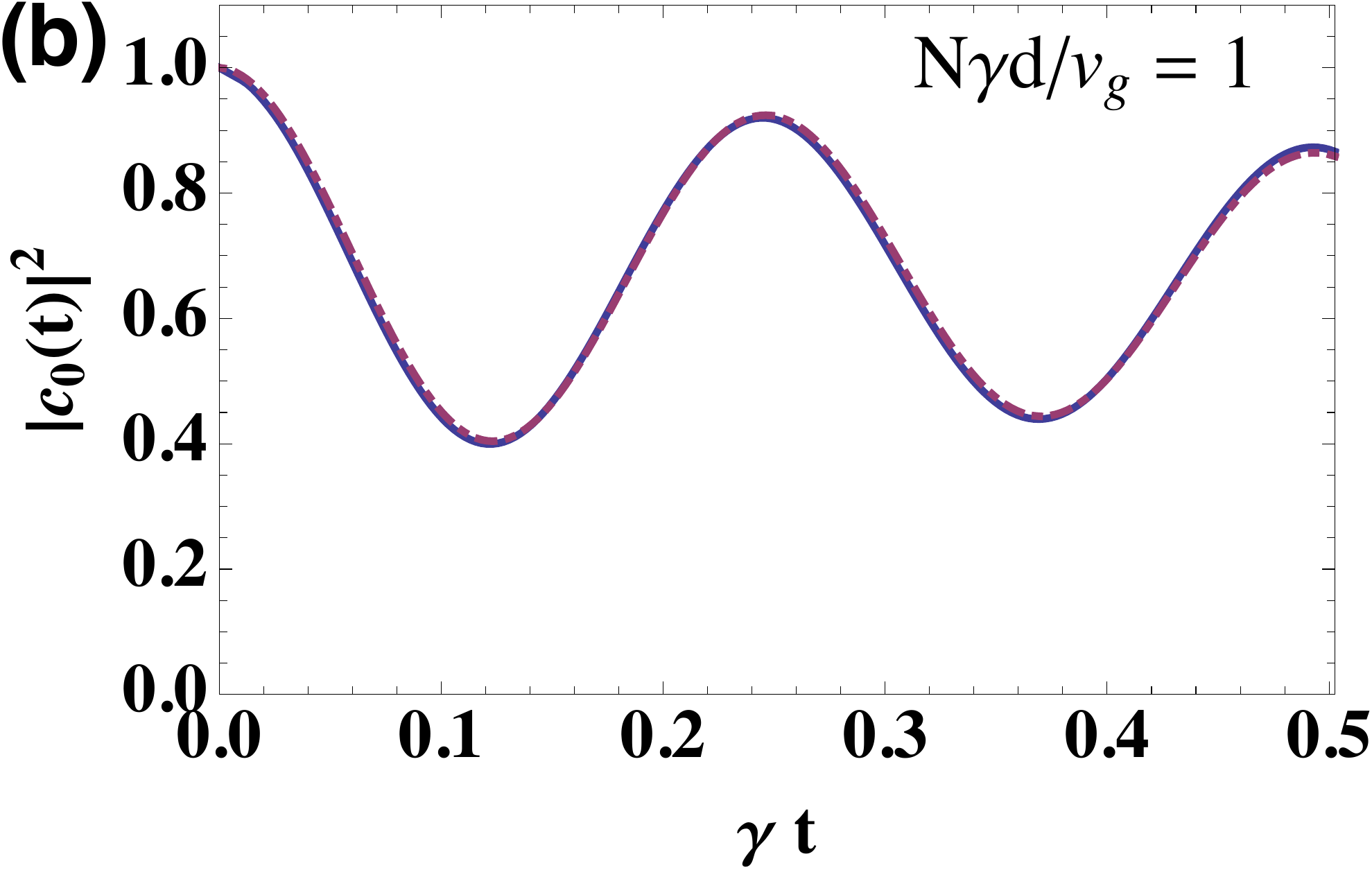}\label{fig:App1}
}
\subfloat{
\includegraphics[width=0.3\textwidth]{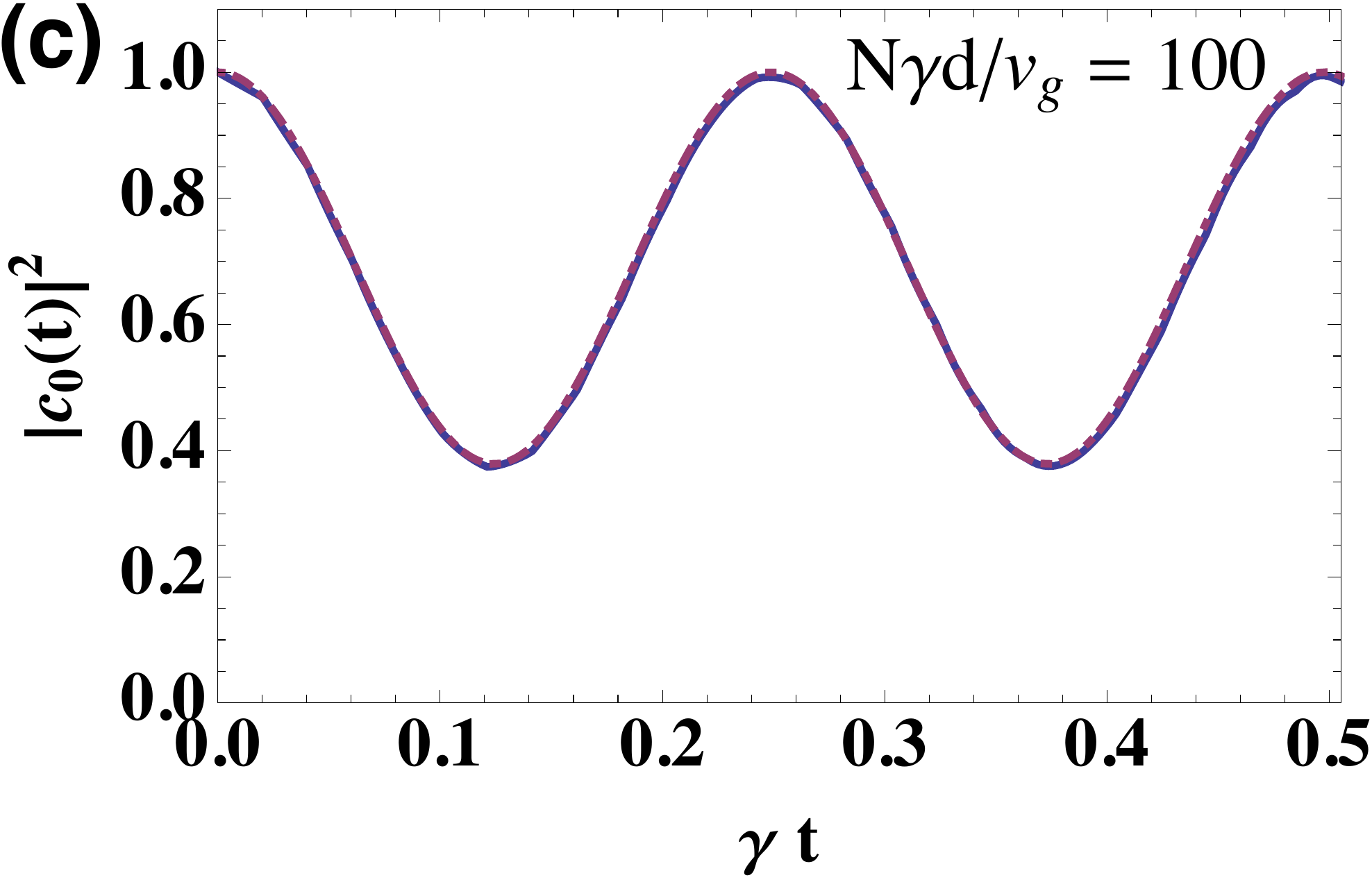}\label{fig:App100}
}
\caption{\label{fig:detune}(color online). Evolution of the excitation of the central atom, for different regimes, with $\gamma d/v_g \ll 1 \ll N$ and $\phi = \pi/10$. The exact solution from Eq.~\eqref{eq:seriesSol} is shown by the solid blue line, and the analytical approximation from Eq.~\eqref{eq:final} the dashed red line. a) $\gamma d/v_g = 0.0002$. b) $\gamma d/v_g = 0.01$. c) $\gamma d/v_g=0.02$.}
\end{figure}

For the general case of arbitrary detuning and arbitrary value of the parameter $a=N \gamma d/v_g$, we also found an analytical approximation that matches the exact formula in Eq.~\eqref{eq:seriesSol} to a high accuracy when $\gamma d/v_g \ll 1 \ll N$, which is given by
\begin{equation}\label{eq:final}\begin{aligned} c_0(t) &= \exp \left(-\frac{\gamma  t}{2 (a+1)^2}\right) \exp \left(\frac{ 2a (a+1)v  }{3 \gamma  d/v_g}\gamma  t\right) \Bigg[\cos \left( \Omega _0 \sqrt{1+u+v}t\right) \\&+\sqrt{\frac{-u}{1+u+v}} \sin \left( \Omega _0 \sqrt{1+u+v}t\right)\Bigg]+\mathcal O(\gamma d/v_g) \end{aligned}
\end{equation} 
with $\Omega_0 = \gamma  \sqrt{\frac{2 a}{(a+1) \gamma d/c}}$, $u = -\frac{2 a (1+a)^3  v^2}{9 \gamma d/v_g}$, and $v = \frac{3}{4} \left(-1+e^{\frac{i \phi }{(1+a)^2}}\right)$. 
The strong agreement between this formula and the exact series solution for different values of $a$ corresponding to the Markovian, non-Markovian and transition regimes can be seen in Fig.~\ref{fig:detune}. 

\end{appendix}
\bibliography{Bibliography}

\begin{thebibliography}{33}%
\makeatletter
\providecommand \@ifxundefined [1]{%
 \@ifx{#1\undefined}
}%
\providecommand \@ifnum [1]{%
 \ifnum #1\expandafter \@firstoftwo
 \else \expandafter \@secondoftwo
 \fi
}%
\providecommand \@ifx [1]{%
 \ifx #1\expandafter \@firstoftwo
 \else \expandafter \@secondoftwo
 \fi
}%
\providecommand \natexlab [1]{#1}%
\providecommand \enquote  [1]{``#1''}%
\providecommand \bibnamefont  [1]{#1}%
\providecommand \bibfnamefont [1]{#1}%
\providecommand \citenamefont [1]{#1}%
\providecommand \href@noop [0]{\@secondoftwo}%
\providecommand \href [0]{\begingroup \@sanitize@url \@href}%
\providecommand \@href[1]{\@@startlink{#1}\@@href}%
\providecommand \@@href[1]{\endgroup#1\@@endlink}%
\providecommand \@sanitize@url [0]{\catcode `\\12\catcode `\$12\catcode
  `\&12\catcode `\#12\catcode `\^12\catcode `\_12\catcode `\%12\relax}%
\providecommand \@@startlink[1]{}%
\providecommand \@@endlink[0]{}%
\providecommand \url  [0]{\begingroup\@sanitize@url \@url }%
\providecommand \@url [1]{\endgroup\@href {#1}{\urlprefix }}%
\providecommand \urlprefix  [0]{URL }%
\providecommand \Eprint [0]{\href }%
\providecommand \doibase [0]{http://dx.doi.org/}%
\providecommand \selectlanguage [0]{\@gobble}%
\providecommand \bibinfo  [0]{\@secondoftwo}%
\providecommand \bibfield  [0]{\@secondoftwo}%
\providecommand \translation [1]{[#1]}%
\providecommand \BibitemOpen [0]{}%
\providecommand \bibitemStop [0]{}%
\providecommand \bibitemNoStop [0]{.\EOS\space}%
\providecommand \EOS [0]{\spacefactor3000\relax}%
\providecommand \BibitemShut  [1]{\csname bibitem#1\endcsname}%
\let\auto@bib@innerbib\@empty
\bibitem [{\citenamefont {Kok}\ and\ \citenamefont
  {Lovett}(2001)}]{opticalqip2011}%
  \BibitemOpen
  \bibfield  {author} {\bibinfo {author} {\bibfnamefont {P.}~\bibnamefont
  {Kok}}\ and\ \bibinfo {author} {\bibfnamefont {B.}~\bibnamefont {Lovett}},\
  }\href@noop {} {\emph {\bibinfo {title} {Introduction to Optical Quantum
  Information Processing}}}\ (\bibinfo  {publisher} {Cambridge University
  Press},\ \bibinfo {year} {2001})\BibitemShut {NoStop}%
\bibitem [{\citenamefont {Turchette}\ \emph {et~al.}(1995)\citenamefont
  {Turchette}, \citenamefont {Hood}, \citenamefont {Lange}, \citenamefont
  {Mabuchi},\ and\ \citenamefont {Kimble}}]{Turchette1995}%
  \BibitemOpen
  \bibfield  {author} {\bibinfo {author} {\bibfnamefont {Q.~A.}\ \bibnamefont
  {Turchette}}, \bibinfo {author} {\bibfnamefont {C.~J.}\ \bibnamefont {Hood}},
  \bibinfo {author} {\bibfnamefont {W.}~\bibnamefont {Lange}}, \bibinfo
  {author} {\bibfnamefont {H.}~\bibnamefont {Mabuchi}}, \ and\ \bibinfo
  {author} {\bibfnamefont {H.~J.}\ \bibnamefont {Kimble}},\ }\href {\doibase
  10.1103/PhysRevLett.75.4710} {\bibfield  {journal} {\bibinfo  {journal}
  {Phys. Rev. Lett.}\ }\textbf {\bibinfo {volume} {75}},\ \bibinfo {pages}
  {4710} (\bibinfo {year} {1995})}\BibitemShut {NoStop}%
\bibitem [{\citenamefont {Domokos}\ \emph {et~al.}(1995)\citenamefont
  {Domokos}, \citenamefont {Raimond}, \citenamefont {Brune},\ and\
  \citenamefont {Haroche}}]{Domokos1995}%
  \BibitemOpen
  \bibfield  {author} {\bibinfo {author} {\bibfnamefont {P.}~\bibnamefont
  {Domokos}}, \bibinfo {author} {\bibfnamefont {J.~M.}\ \bibnamefont
  {Raimond}}, \bibinfo {author} {\bibfnamefont {M.}~\bibnamefont {Brune}}, \
  and\ \bibinfo {author} {\bibfnamefont {S.}~\bibnamefont {Haroche}},\ }\href
  {\doibase 10.1103/PhysRevA.52.3554} {\bibfield  {journal} {\bibinfo
  {journal} {Phys. Rev. A}\ }\textbf {\bibinfo {volume} {52}},\ \bibinfo
  {pages} {3554} (\bibinfo {year} {1995})}\BibitemShut {NoStop}%
\bibitem [{\citenamefont {Imamo\u{g}lu}\ \emph {et~al.}(1999)\citenamefont
  {Imamo\u{g}lu}, \citenamefont {Awschalom}, \citenamefont {Burkard},
  \citenamefont {DiVincenzo}, \citenamefont {Loss}, \citenamefont {Sherwin},\
  and\ \citenamefont {Small}}]{Imamoglu1999}%
  \BibitemOpen
  \bibfield  {author} {\bibinfo {author} {\bibfnamefont {A.}~\bibnamefont
  {Imamo\u{g}lu}}, \bibinfo {author} {\bibfnamefont {D.~D.}\ \bibnamefont
  {Awschalom}}, \bibinfo {author} {\bibfnamefont {G.}~\bibnamefont {Burkard}},
  \bibinfo {author} {\bibfnamefont {D.~P.}\ \bibnamefont {DiVincenzo}},
  \bibinfo {author} {\bibfnamefont {D.}~\bibnamefont {Loss}}, \bibinfo {author}
  {\bibfnamefont {M.}~\bibnamefont {Sherwin}}, \ and\ \bibinfo {author}
  {\bibfnamefont {A.}~\bibnamefont {Small}},\ }\href {\doibase
  10.1103/PhysRevLett.83.4204} {\bibfield  {journal} {\bibinfo  {journal}
  {Phys. Rev. Lett.}\ }\textbf {\bibinfo {volume} {83}},\ \bibinfo {pages}
  {4204} (\bibinfo {year} {1999})}\BibitemShut {NoStop}%
\bibitem [{\citenamefont {Zheng}\ and\ \citenamefont {Guo}(2000)}]{Zheng2000}%
  \BibitemOpen
  \bibfield  {author} {\bibinfo {author} {\bibfnamefont {S.-B.}\ \bibnamefont
  {Zheng}}\ and\ \bibinfo {author} {\bibfnamefont {G.-C.}\ \bibnamefont
  {Guo}},\ }\href {\doibase 10.1103/PhysRevLett.85.2392} {\bibfield  {journal}
  {\bibinfo  {journal} {Phys. Rev. Lett.}\ }\textbf {\bibinfo {volume} {85}},\
  \bibinfo {pages} {2392} (\bibinfo {year} {2000})}\BibitemShut {NoStop}%
\bibitem [{\citenamefont {Blais}\ \emph {et~al.}(2004)\citenamefont {Blais},
  \citenamefont {Huang}, \citenamefont {Wallraff}, \citenamefont {Girvin},\
  and\ \citenamefont {Schoelkopf}}]{Blais2004}%
  \BibitemOpen
  \bibfield  {author} {\bibinfo {author} {\bibfnamefont {A.}~\bibnamefont
  {Blais}}, \bibinfo {author} {\bibfnamefont {R.-S.}\ \bibnamefont {Huang}},
  \bibinfo {author} {\bibfnamefont {A.}~\bibnamefont {Wallraff}}, \bibinfo
  {author} {\bibfnamefont {S.~M.}\ \bibnamefont {Girvin}}, \ and\ \bibinfo
  {author} {\bibfnamefont {R.~J.}\ \bibnamefont {Schoelkopf}},\ }\href
  {\doibase 10.1103/PhysRevA.69.062320} {\bibfield  {journal} {\bibinfo
  {journal} {Phys. Rev. A}\ }\textbf {\bibinfo {volume} {69}},\ \bibinfo
  {pages} {062320} (\bibinfo {year} {2004})}\BibitemShut {NoStop}%
\bibitem [{\citenamefont {Christensen}\ \emph {et~al.}(2008)\citenamefont
  {Christensen}, \citenamefont {Will}, \citenamefont {Saba}, \citenamefont
  {Jo}, \citenamefont {Shin}, \citenamefont {Ketterle},\ and\ \citenamefont
  {Pritchard}}]{Christensen2008}%
  \BibitemOpen
  \bibfield  {author} {\bibinfo {author} {\bibfnamefont {C.~A.}\ \bibnamefont
  {Christensen}}, \bibinfo {author} {\bibfnamefont {S.}~\bibnamefont {Will}},
  \bibinfo {author} {\bibfnamefont {M.}~\bibnamefont {Saba}}, \bibinfo {author}
  {\bibfnamefont {G.-B.}\ \bibnamefont {Jo}}, \bibinfo {author} {\bibfnamefont
  {Y.-I.}\ \bibnamefont {Shin}}, \bibinfo {author} {\bibfnamefont
  {W.}~\bibnamefont {Ketterle}}, \ and\ \bibinfo {author} {\bibfnamefont
  {D.}~\bibnamefont {Pritchard}},\ }\href {\doibase 10.1103/PhysRevA.78.033429}
  {\bibfield  {journal} {\bibinfo  {journal} {Phys. Rev. A}\ }\textbf {\bibinfo
  {volume} {78}},\ \bibinfo {pages} {033429} (\bibinfo {year}
  {2008})}\BibitemShut {NoStop}%
\bibitem [{\citenamefont {Bajcsy}\ \emph {et~al.}(2011)\citenamefont {Bajcsy},
  \citenamefont {Hofferberth}, \citenamefont {Peyronel}, \citenamefont {Balic},
  \citenamefont {Liang}, \citenamefont {Zibrov}, \citenamefont {Vuletic},\ and\
  \citenamefont {Lukin}}]{Bajcsy2011}%
  \BibitemOpen
  \bibfield  {author} {\bibinfo {author} {\bibfnamefont {M.}~\bibnamefont
  {Bajcsy}}, \bibinfo {author} {\bibfnamefont {S.}~\bibnamefont {Hofferberth}},
  \bibinfo {author} {\bibfnamefont {T.}~\bibnamefont {Peyronel}}, \bibinfo
  {author} {\bibfnamefont {V.}~\bibnamefont {Balic}}, \bibinfo {author}
  {\bibfnamefont {Q.}~\bibnamefont {Liang}}, \bibinfo {author} {\bibfnamefont
  {A.~S.}\ \bibnamefont {Zibrov}}, \bibinfo {author} {\bibfnamefont
  {V.}~\bibnamefont {Vuletic}}, \ and\ \bibinfo {author} {\bibfnamefont
  {M.~D.}\ \bibnamefont {Lukin}},\ }\href {\doibase 10.1103/PhysRevA.83.063830}
  {\bibfield  {journal} {\bibinfo  {journal} {Phys. Rev. A}\ }\textbf {\bibinfo
  {volume} {83}},\ \bibinfo {pages} {063830} (\bibinfo {year}
  {2011})}\BibitemShut {NoStop}%
\bibitem [{\citenamefont {Vetsch}\ \emph {et~al.}(2010)\citenamefont {Vetsch},
  \citenamefont {Reitz}, \citenamefont {Sagu\'e}, \citenamefont {Schmidt},
  \citenamefont {Dawkins},\ and\ \citenamefont {Rauschenbeutel}}]{Vetsch2010}%
  \BibitemOpen
  \bibfield  {author} {\bibinfo {author} {\bibfnamefont {E.}~\bibnamefont
  {Vetsch}}, \bibinfo {author} {\bibfnamefont {D.}~\bibnamefont {Reitz}},
  \bibinfo {author} {\bibfnamefont {G.}~\bibnamefont {Sagu\'e}}, \bibinfo
  {author} {\bibfnamefont {R.}~\bibnamefont {Schmidt}}, \bibinfo {author}
  {\bibfnamefont {S.~T.}\ \bibnamefont {Dawkins}}, \ and\ \bibinfo {author}
  {\bibfnamefont {A.}~\bibnamefont {Rauschenbeutel}},\ }\href {\doibase
  10.1103/PhysRevLett.104.203603} {\bibfield  {journal} {\bibinfo  {journal}
  {Phys. Rev. Lett.}\ }\textbf {\bibinfo {volume} {104}},\ \bibinfo {pages}
  {203603} (\bibinfo {year} {2010})}\BibitemShut {NoStop}%
\bibitem [{\citenamefont {Goban}\ \emph {et~al.}(2012)\citenamefont {Goban},
  \citenamefont {Choi}, \citenamefont {Alton}, \citenamefont {Ding},
  \citenamefont {Lacro\^ute}, \citenamefont {Pototschnig}, \citenamefont
  {Thiele}, \citenamefont {Stern},\ and\ \citenamefont {Kimble}}]{Goban2012}%
  \BibitemOpen
  \bibfield  {author} {\bibinfo {author} {\bibfnamefont {A.}~\bibnamefont
  {Goban}}, \bibinfo {author} {\bibfnamefont {K.~S.}\ \bibnamefont {Choi}},
  \bibinfo {author} {\bibfnamefont {D.~J.}\ \bibnamefont {Alton}}, \bibinfo
  {author} {\bibfnamefont {D.}~\bibnamefont {Ding}}, \bibinfo {author}
  {\bibfnamefont {C.}~\bibnamefont {Lacro\^ute}}, \bibinfo {author}
  {\bibfnamefont {M.}~\bibnamefont {Pototschnig}}, \bibinfo {author}
  {\bibfnamefont {T.}~\bibnamefont {Thiele}}, \bibinfo {author} {\bibfnamefont
  {N.~P.}\ \bibnamefont {Stern}}, \ and\ \bibinfo {author} {\bibfnamefont
  {H.~J.}\ \bibnamefont {Kimble}},\ }\href {\doibase
  10.1103/PhysRevLett.109.033603} {\bibfield  {journal} {\bibinfo  {journal}
  {Phys. Rev. Lett.}\ }\textbf {\bibinfo {volume} {109}},\ \bibinfo {pages}
  {033603} (\bibinfo {year} {2012})}\BibitemShut {NoStop}%
\bibitem [{\citenamefont {Arcari}\ \emph {et~al.}(2014)\citenamefont {Arcari},
  \citenamefont {S\"ollner}, \citenamefont {Javadi}, \citenamefont
  {Lindskov~Hansen}, \citenamefont {Mahmoodian}, \citenamefont {Liu},
  \citenamefont {Thyrrestrup}, \citenamefont {Lee}, \citenamefont {Song},
  \citenamefont {Stobbe},\ and\ \citenamefont {Lodahl}}]{Arcari2014}%
  \BibitemOpen
  \bibfield  {author} {\bibinfo {author} {\bibfnamefont {M.}~\bibnamefont
  {Arcari}}, \bibinfo {author} {\bibfnamefont {I.}~\bibnamefont {S\"ollner}},
  \bibinfo {author} {\bibfnamefont {A.}~\bibnamefont {Javadi}}, \bibinfo
  {author} {\bibfnamefont {S.}~\bibnamefont {Lindskov~Hansen}}, \bibinfo
  {author} {\bibfnamefont {S.}~\bibnamefont {Mahmoodian}}, \bibinfo {author}
  {\bibfnamefont {J.}~\bibnamefont {Liu}}, \bibinfo {author} {\bibfnamefont
  {H.}~\bibnamefont {Thyrrestrup}}, \bibinfo {author} {\bibfnamefont
  {E.}~\bibnamefont {Lee}}, \bibinfo {author} {\bibfnamefont {J.}~\bibnamefont
  {Song}}, \bibinfo {author} {\bibfnamefont {S.}~\bibnamefont {Stobbe}}, \ and\
  \bibinfo {author} {\bibfnamefont {P.}~\bibnamefont {Lodahl}},\ }\href
  {\doibase 10.1103/PhysRevLett.113.093603} {\bibfield  {journal} {\bibinfo
  {journal} {Phys. Rev. Lett.}\ }\textbf {\bibinfo {volume} {113}},\ \bibinfo
  {pages} {093603} (\bibinfo {year} {2014})}\BibitemShut {NoStop}%
\bibitem [{\citenamefont {Goban}\ \emph {et~al.}(2014)\citenamefont {Goban},
  \citenamefont {Hung}, \citenamefont {Hood}, \citenamefont {Muniz},
  \citenamefont {Lee}, \citenamefont {Martin}, \citenamefont {McClung},
  \citenamefont {Choi}, \citenamefont {Chang}, \citenamefont {Painter},\ and\
  \citenamefont {Kimble}}]{Goban2014}%
  \BibitemOpen
  \bibfield  {author} {\bibinfo {author} {\bibfnamefont {A.}~\bibnamefont
  {Goban}}, \bibinfo {author} {\bibfnamefont {S.-P.}\ \bibnamefont {Hung},
  \bibfnamefont {C.-L.~Yu}}, \bibinfo {author} {\bibfnamefont {J.}~\bibnamefont
  {Hood}}, \bibinfo {author} {\bibfnamefont {J.}~\bibnamefont {Muniz}},
  \bibinfo {author} {\bibfnamefont {J.}~\bibnamefont {Lee}}, \bibinfo {author}
  {\bibfnamefont {M.}~\bibnamefont {Martin}}, \bibinfo {author} {\bibfnamefont
  {A.}~\bibnamefont {McClung}}, \bibinfo {author} {\bibfnamefont
  {K.}~\bibnamefont {Choi}}, \bibinfo {author} {\bibfnamefont {D.}~\bibnamefont
  {Chang}}, \bibinfo {author} {\bibfnamefont {O.}~\bibnamefont {Painter}}, \
  and\ \bibinfo {author} {\bibfnamefont {H.}~\bibnamefont {Kimble}},\ }\href
  {\doibase http://dx.doi.org/10.1038/ncomms4808} {\bibfield  {journal}
  {\bibinfo  {journal} {Nature Communications}\ }\textbf {\bibinfo {volume}
  {5}},\ \bibinfo {pages} {3808} (\bibinfo {year} {2014})}\BibitemShut
  {NoStop}%
\bibitem [{\citenamefont {Chang}\ \emph {et~al.}(2012)\citenamefont {Chang},
  \citenamefont {Jiang}, \citenamefont {Gorshkov},\ and\ \citenamefont
  {Kimble}}]{Chang2012}%
  \BibitemOpen
  \bibfield  {author} {\bibinfo {author} {\bibfnamefont {D.~E.}\ \bibnamefont
  {Chang}}, \bibinfo {author} {\bibfnamefont {L.}~\bibnamefont {Jiang}},
  \bibinfo {author} {\bibfnamefont {A.~V.}\ \bibnamefont {Gorshkov}}, \ and\
  \bibinfo {author} {\bibfnamefont {H.~J.}\ \bibnamefont {Kimble}},\ }\href
  {http://stacks.iop.org/1367-2630/14/i=6/a=063003} {\bibfield  {journal}
  {\bibinfo  {journal} {New Journal of Physics}\ }\textbf {\bibinfo {volume}
  {14}},\ \bibinfo {pages} {063003} (\bibinfo {year} {2012})}\BibitemShut
  {NoStop}%
\bibitem [{\citenamefont {Zheng}\ and\ \citenamefont
  {Baranger}(2013)}]{Zheng2013}%
  \BibitemOpen
  \bibfield  {author} {\bibinfo {author} {\bibfnamefont {H.}~\bibnamefont
  {Zheng}}\ and\ \bibinfo {author} {\bibfnamefont {H.~U.}\ \bibnamefont
  {Baranger}},\ }\href {\doibase 10.1103/PhysRevLett.110.113601} {\bibfield
  {journal} {\bibinfo  {journal} {Phys. Rev. Lett.}\ }\textbf {\bibinfo
  {volume} {110}},\ \bibinfo {pages} {113601} (\bibinfo {year}
  {2013})}\BibitemShut {NoStop}%
\bibitem [{\citenamefont {Shi}\ \emph {et~al.}(2015)\citenamefont {Shi},
  \citenamefont {Chang},\ and\ \citenamefont {Cirac}}]{Shi2015}%
  \BibitemOpen
  \bibfield  {author} {\bibinfo {author} {\bibfnamefont {T.}~\bibnamefont
  {Shi}}, \bibinfo {author} {\bibfnamefont {D.~E.}\ \bibnamefont {Chang}}, \
  and\ \bibinfo {author} {\bibfnamefont {J.~I.}\ \bibnamefont {Cirac}},\ }\href
  {\doibase 10.1103/PhysRevA.92.053834} {\bibfield  {journal} {\bibinfo
  {journal} {Phys. Rev. A}\ }\textbf {\bibinfo {volume} {92}},\ \bibinfo
  {pages} {053834} (\bibinfo {year} {2015})}\BibitemShut {NoStop}%
\bibitem [{\citenamefont {Tsoi}\ and\ \citenamefont {Law}(2008)}]{Tsoi2008}%
  \BibitemOpen
  \bibfield  {author} {\bibinfo {author} {\bibfnamefont {T.~S.}\ \bibnamefont
  {Tsoi}}\ and\ \bibinfo {author} {\bibfnamefont {C.~K.}\ \bibnamefont {Law}},\
  }\href {\doibase 10.1103/PhysRevA.78.063832} {\bibfield  {journal} {\bibinfo
  {journal} {Phys. Rev. A}\ }\textbf {\bibinfo {volume} {78}},\ \bibinfo
  {pages} {063832} (\bibinfo {year} {2008})}\BibitemShut {NoStop}%
\bibitem [{\citenamefont {Birkl}\ \emph {et~al.}(1995)\citenamefont {Birkl},
  \citenamefont {Gatzke}, \citenamefont {Deutsch}, \citenamefont {Rolston},\
  and\ \citenamefont {Phillips}}]{Birkl1995}%
  \BibitemOpen
  \bibfield  {author} {\bibinfo {author} {\bibfnamefont {G.}~\bibnamefont
  {Birkl}}, \bibinfo {author} {\bibfnamefont {M.}~\bibnamefont {Gatzke}},
  \bibinfo {author} {\bibfnamefont {I.~H.}\ \bibnamefont {Deutsch}}, \bibinfo
  {author} {\bibfnamefont {S.~L.}\ \bibnamefont {Rolston}}, \ and\ \bibinfo
  {author} {\bibfnamefont {W.~D.}\ \bibnamefont {Phillips}},\ }\href {\doibase
  10.1103/PhysRevLett.75.2823} {\bibfield  {journal} {\bibinfo  {journal}
  {Phys. Rev. Lett.}\ }\textbf {\bibinfo {volume} {75}},\ \bibinfo {pages}
  {2823} (\bibinfo {year} {1995})}\BibitemShut {NoStop}%
\bibitem [{\citenamefont {Schilke}\ \emph {et~al.}(2011)\citenamefont
  {Schilke}, \citenamefont {Zimmermann}, \citenamefont {Courteille},\ and\
  \citenamefont {Guerin}}]{Schilke2011}%
  \BibitemOpen
  \bibfield  {author} {\bibinfo {author} {\bibfnamefont {A.}~\bibnamefont
  {Schilke}}, \bibinfo {author} {\bibfnamefont {C.}~\bibnamefont {Zimmermann}},
  \bibinfo {author} {\bibfnamefont {P.~W.}\ \bibnamefont {Courteille}}, \ and\
  \bibinfo {author} {\bibfnamefont {W.}~\bibnamefont {Guerin}},\ }\href
  {\doibase 10.1103/PhysRevLett.106.223903} {\bibfield  {journal} {\bibinfo
  {journal} {Phys. Rev. Lett.}\ }\textbf {\bibinfo {volume} {106}},\ \bibinfo
  {pages} {223903} (\bibinfo {year} {2011})}\BibitemShut {NoStop}%
\bibitem [{\citenamefont {{Goban}}\ \emph {et~al.}(2014)\citenamefont
  {{Goban}}, \citenamefont {{Hung}}, \citenamefont {{Hood}}, \citenamefont
  {{Yu}}, \citenamefont {{Muniz}}, \citenamefont {{Painter}},\ and\
  \citenamefont {{Kimble}}}]{Goban2015}%
  \BibitemOpen
  \bibfield  {author} {\bibinfo {author} {\bibfnamefont {A.}~\bibnamefont
  {{Goban}}}, \bibinfo {author} {\bibfnamefont {C.~L.}\ \bibnamefont {{Hung}}},
  \bibinfo {author} {\bibfnamefont {J.~D.}\ \bibnamefont {{Hood}}}, \bibinfo
  {author} {\bibfnamefont {S.~P.}\ \bibnamefont {{Yu}}}, \bibinfo {author}
  {\bibfnamefont {J.~A.}\ \bibnamefont {{Muniz}}}, \bibinfo {author}
  {\bibfnamefont {O.}~\bibnamefont {{Painter}}}, \ and\ \bibinfo {author}
  {\bibfnamefont {H.~J.}\ \bibnamefont {{Kimble}}},\ }\href@noop {} {\
  (\bibinfo {year} {2014})},\ \Eprint {http://arxiv.org/abs/1503.04503}
  {arXiv:1503.04503 [physics.optics]} \BibitemShut {NoStop}%
\bibitem [{\citenamefont {Domokos}\ \emph {et~al.}(2002)\citenamefont
  {Domokos}, \citenamefont {Horak},\ and\ \citenamefont
  {Ritsch}}]{Domokos2002}%
  \BibitemOpen
  \bibfield  {author} {\bibinfo {author} {\bibfnamefont {P.}~\bibnamefont
  {Domokos}}, \bibinfo {author} {\bibfnamefont {P.}~\bibnamefont {Horak}}, \
  and\ \bibinfo {author} {\bibfnamefont {H.}~\bibnamefont {Ritsch}},\ }\href
  {\doibase 10.1103/PhysRevA.65.033832} {\bibfield  {journal} {\bibinfo
  {journal} {Phys. Rev. A}\ }\textbf {\bibinfo {volume} {65}},\ \bibinfo
  {pages} {033832} (\bibinfo {year} {2002})}\BibitemShut {NoStop}%
\bibitem [{\citenamefont {Hoi}\ \emph {et~al.}(2011)\citenamefont {Hoi},
  \citenamefont {Wilson}, \citenamefont {Johansson}, \citenamefont {Palomaki},
  \citenamefont {Peropadre},\ and\ \citenamefont {Delsing}}]{Hoi2011}%
  \BibitemOpen
  \bibfield  {author} {\bibinfo {author} {\bibfnamefont {I.-C.}\ \bibnamefont
  {Hoi}}, \bibinfo {author} {\bibfnamefont {C.~M.}\ \bibnamefont {Wilson}},
  \bibinfo {author} {\bibfnamefont {G.}~\bibnamefont {Johansson}}, \bibinfo
  {author} {\bibfnamefont {T.}~\bibnamefont {Palomaki}}, \bibinfo {author}
  {\bibfnamefont {B.}~\bibnamefont {Peropadre}}, \ and\ \bibinfo {author}
  {\bibfnamefont {P.}~\bibnamefont {Delsing}},\ }\href {\doibase
  10.1103/PhysRevLett.107.073601} {\bibfield  {journal} {\bibinfo  {journal}
  {Phys. Rev. Lett.}\ }\textbf {\bibinfo {volume} {107}},\ \bibinfo {pages}
  {073601} (\bibinfo {year} {2011})}\BibitemShut {NoStop}%
\bibitem [{\citenamefont {Cohen-Tannoudji}\ \emph {et~al.}(1998)\citenamefont
  {Cohen-Tannoudji}, \citenamefont {Dupont-Roc},\ and\ \citenamefont
  {Grynberg}}]{atphint1998}%
  \BibitemOpen
  \bibfield  {author} {\bibinfo {author} {\bibfnamefont {C.}~\bibnamefont
  {Cohen-Tannoudji}}, \bibinfo {author} {\bibfnamefont {J.}~\bibnamefont
  {Dupont-Roc}}, \ and\ \bibinfo {author} {\bibfnamefont {G.}~\bibnamefont
  {Grynberg}},\ }\href@noop {} {\emph {\bibinfo {title} {Atom-Photon
  Interactions}}}\ (\bibinfo  {publisher} {Wiley-VCH},\ \bibinfo {year}
  {1998})\BibitemShut {NoStop}%
\bibitem [{\citenamefont {Shen}\ and\ \citenamefont {Fan}(2005)}]{SFopt2005}%
  \BibitemOpen
  \bibfield  {author} {\bibinfo {author} {\bibfnamefont {J.~T.}\ \bibnamefont
  {Shen}}\ and\ \bibinfo {author} {\bibfnamefont {S.}~\bibnamefont {Fan}},\
  }\href {\doibase 10.1364/OL.30.002001} {\bibfield  {journal} {\bibinfo
  {journal} {Opt. Lett.}\ }\textbf {\bibinfo {volume} {30}},\ \bibinfo {pages}
  {2001} (\bibinfo {year} {2005})}\BibitemShut {NoStop}%
\bibitem [{\citenamefont {Barnett}\ and\ \citenamefont
  {Radmore}(2002)}]{methodsQO2002}%
  \BibitemOpen
  \bibfield  {author} {\bibinfo {author} {\bibfnamefont {S.}~\bibnamefont
  {Barnett}}\ and\ \bibinfo {author} {\bibfnamefont {P.}~\bibnamefont
  {Radmore}},\ }\href@noop {} {\emph {\bibinfo {title} {Methods in Theoretical
  Quantum Optics}}}\ (\bibinfo  {publisher} {Oxford University Press},\
  \bibinfo {year} {2002})\BibitemShut {NoStop}%
\bibitem [{\citenamefont {Milonni}\ and\ \citenamefont
  {Knight}(1974)}]{Milonni1974}%
  \BibitemOpen
  \bibfield  {author} {\bibinfo {author} {\bibfnamefont {P.~W.}\ \bibnamefont
  {Milonni}}\ and\ \bibinfo {author} {\bibfnamefont {P.~L.}\ \bibnamefont
  {Knight}},\ }\href {\doibase 10.1103/PhysRevA.10.1096} {\bibfield  {journal}
  {\bibinfo  {journal} {Phys. Rev. A}\ }\textbf {\bibinfo {volume} {10}},\
  \bibinfo {pages} {1096} (\bibinfo {year} {1974})}\BibitemShut {NoStop}%
\bibitem [{\citenamefont {Dorner}\ and\ \citenamefont
  {Zoller}(2002)}]{Dorner2002}%
  \BibitemOpen
  \bibfield  {author} {\bibinfo {author} {\bibfnamefont {U.}~\bibnamefont
  {Dorner}}\ and\ \bibinfo {author} {\bibfnamefont {P.}~\bibnamefont
  {Zoller}},\ }\href {\doibase 10.1103/PhysRevA.66.023816} {\bibfield
  {journal} {\bibinfo  {journal} {Phys. Rev. A}\ }\textbf {\bibinfo {volume}
  {66}},\ \bibinfo {pages} {023816} (\bibinfo {year} {2002})}\BibitemShut
  {NoStop}%
\bibitem [{\citenamefont {Hood}\ \emph {et~al.}(2001)\citenamefont {Hood},
  \citenamefont {Kimble},\ and\ \citenamefont {Ye}}]{Kimble2001}%
  \BibitemOpen
  \bibfield  {author} {\bibinfo {author} {\bibfnamefont {C.~J.}\ \bibnamefont
  {Hood}}, \bibinfo {author} {\bibfnamefont {H.~J.}\ \bibnamefont {Kimble}}, \
  and\ \bibinfo {author} {\bibfnamefont {J.}~\bibnamefont {Ye}},\ }\href
  {\doibase 10.1103/PhysRevA.64.033804} {\bibfield  {journal} {\bibinfo
  {journal} {Phys. Rev. A}\ }\textbf {\bibinfo {volume} {64}},\ \bibinfo
  {pages} {033804} (\bibinfo {year} {2001})}\BibitemShut {NoStop}%
\bibitem [{\citenamefont {Baba}(2008)}]{Baba2008}%
  \BibitemOpen
  \bibfield  {author} {\bibinfo {author} {\bibfnamefont {T.}~\bibnamefont
  {Baba}},\ }\href {http://dx.doi.org/10.1038/nphoton.2008.146} {\bibfield
  {journal} {\bibinfo  {journal} {Nat Photon}\ }\textbf {\bibinfo {volume}
  {2}},\ \bibinfo {pages} {465} (\bibinfo {year} {2008})}\BibitemShut {NoStop}%
\bibitem [{\citenamefont {Li}\ \emph {et~al.}(2012)\citenamefont {Li},
  \citenamefont {O\'Faolain}, \citenamefont {Schulz},\ and\ \citenamefont
  {Krauss}}]{Li2012}%
  \BibitemOpen
  \bibfield  {author} {\bibinfo {author} {\bibfnamefont {J.}~\bibnamefont
  {Li}}, \bibinfo {author} {\bibfnamefont {L.}~\bibnamefont {O\'Faolain}},
  \bibinfo {author} {\bibfnamefont {S.~A.}\ \bibnamefont {Schulz}}, \ and\
  \bibinfo {author} {\bibfnamefont {T.~F.}\ \bibnamefont {Krauss}},\ }\href
  {\doibase http://dx.doi.org/10.1016/j.photonics.2012.05.006} {\bibfield
  {journal} {\bibinfo  {journal} {Photonics and Nanostructures - Fundamentals
  and Applications}\ }\textbf {\bibinfo {volume} {10}},\ \bibinfo {pages} {589
  } (\bibinfo {year} {2012})}\BibitemShut {NoStop}%
\bibitem [{\citenamefont {Agarwal}(2012)}]{Agarwal2012}%
  \BibitemOpen
  \bibfield  {author} {\bibinfo {author} {\bibfnamefont {G.}~\bibnamefont
  {Agarwal}},\ }\href@noop {} {\emph {\bibinfo {title} {Quantum Optics}}}\
  (\bibinfo  {publisher} {Cambridge University Press},\ \bibinfo {year}
  {2012})\BibitemShut {NoStop}%
\bibitem [{\citenamefont {Tiecke}\ \emph {et~al.}(2014)\citenamefont {Tiecke},
  \citenamefont {Thompson}, \citenamefont {de~Leon}, \citenamefont {Liu},
  \citenamefont {Vuletic},\ and\ \citenamefont {Lukin}}]{Tiecke2014}%
  \BibitemOpen
  \bibfield  {author} {\bibinfo {author} {\bibfnamefont {T.~G.}\ \bibnamefont
  {Tiecke}}, \bibinfo {author} {\bibfnamefont {J.~D.}\ \bibnamefont
  {Thompson}}, \bibinfo {author} {\bibfnamefont {N.~P.}\ \bibnamefont
  {de~Leon}}, \bibinfo {author} {\bibfnamefont {L.~R.}\ \bibnamefont {Liu}},
  \bibinfo {author} {\bibfnamefont {V.}~\bibnamefont {Vuletic}}, \ and\
  \bibinfo {author} {\bibfnamefont {M.~D.}\ \bibnamefont {Lukin}},\ }\href
  {http://dx.doi.org/10.1038/nature13188} {\bibfield  {journal} {\bibinfo
  {journal} {Nature}\ }\textbf {\bibinfo {volume} {508}},\ \bibinfo {pages}
  {241} (\bibinfo {year} {2014})}\BibitemShut {NoStop}%
\bibitem [{\citenamefont {Reiserer}\ \emph {et~al.}(2014)\citenamefont
  {Reiserer}, \citenamefont {Kalb}, \citenamefont {Rempe},\ and\ \citenamefont
  {Ritter}}]{Reiserer2014}%
  \BibitemOpen
  \bibfield  {author} {\bibinfo {author} {\bibfnamefont {A.}~\bibnamefont
  {Reiserer}}, \bibinfo {author} {\bibfnamefont {N.}~\bibnamefont {Kalb}},
  \bibinfo {author} {\bibfnamefont {G.}~\bibnamefont {Rempe}}, \ and\ \bibinfo
  {author} {\bibfnamefont {S.}~\bibnamefont {Ritter}},\ }\href
  {http://dx.doi.org/10.1038/nature13177} {\bibfield  {journal} {\bibinfo
  {journal} {Nature}\ }\textbf {\bibinfo {volume} {508}},\ \bibinfo {pages}
  {237} (\bibinfo {year} {2014})}\BibitemShut {NoStop}%
\bibitem [{\citenamefont {Parker}\ and\ \citenamefont
  {Stroud}(1987)}]{Parker1987}%
  \BibitemOpen
  \bibfield  {author} {\bibinfo {author} {\bibfnamefont {J.}~\bibnamefont
  {Parker}}\ and\ \bibinfo {author} {\bibfnamefont {C.~R.}\ \bibnamefont
  {Stroud}},\ }\href {\doibase 10.1103/PhysRevA.35.4226} {\bibfield  {journal}
  {\bibinfo  {journal} {Phys. Rev. A}\ }\textbf {\bibinfo {volume} {35}},\
  \bibinfo {pages} {4226} (\bibinfo {year} {1987})}\BibitemShut {NoStop}%
\end{thebibliography}%

\end{document}